\pdfoutput=1% For pdflatex
\documentclass[aps,prc,preprint,nofootinbib,preprintnumbers,superscriptaddress,tightenlines]{revtex4-2}
\usepackage{amsmath,mathrsfs}    % need for subequations
\usepackage{graphicx}   % need for figures
\usepackage{subfigure}
\usepackage{color}
\usepackage[colorlinks=true,linkcolor=blue,citecolor=blue, urlcolor=blue]{hyperref}   % use for hypertext links, including those to external documents and URLs
\usepackage{bm} % for bold caligraphic font
\usepackage{ulem}
\usepackage[inline]{enumitem}
\usepackage{empheq}

\usepackage{soul}
\usepackage{graphicx}
\usepackage{amsmath,amssymb}
\usepackage{soul}
\usepackage{textcomp}
\usepackage{hyperref}
\usepackage{subfigure}
\usepackage[toc,page]{appendix}

\def\p{{\bm p}}

\def\q{|{\bm q}|}
\def\k{{\bm k}}

\def\Tr{{\rm Tr}}

\def\sym{{\rm sym}}

\def\pa{\partial}

\def\k{{\bm k}}

\def\pl{{\parallel}}

\def\st{\begin{equation}}
\def\stp{\end{equation}}
\def\bg{\begin{eqnarray}}
\def\nd{\end{eqnarray}}
\def\Eq#1{Eq.~(\ref{#1})}
\def\Eqs#1{Eqs.~(\ref{#1})}

\def\App#1{App.~\ref{#1}}
\def\Fig#1{Fig.~\ref{#1}}

\def\Sect#1{Sect.~\ref{#1}}
\def\Ref#1{Ref.~\cite{#1}}
\def\llangle{\left\langle}
\def\rrangle{\right\rangle}
\def \bes {\begin{subequations}}
\def \ees {\end{subequations}}

\def\tr{\mathrm{tr}}

\def\tr{{\mathrm{tr}}}
\def\ta{{\mathcal T}_A}

\newcommand{\longi}{\mathbb{L}}
\newcommand{\trans}{\mathbb{T}}
\newcommand{\nn}{\nonumber}
\newcommand{\avg}{{\bar{\gamma}_p}}

\begin{document}

%\preprint{}

%\title{Kinetics of soft pions near the chiral limit}
\title{Kinetics of hydrodynamic pions in chiral perturbation theory}
\author{Juan M. Torres-Rincon}
\email[]{torres@fqa.ub.edu}
\affiliation{Departament de F\'isica Qu\`antica i Astrof\'isica and Institut de Ci\`encies del Cosmos (ICCUB), Facultat de F\'isica,  Universitat de Barcelona, Mart\'i i Franqu\`es 1, 08028 Barcelona, Spain}
\affiliation{Institut f\"ur Theoretische Physik, Goethe Universit\"at Frankfurt, Max von Laue Strasse 1, 60438 Frankfurt, Germany}

\author{Derek Teaney}
\email[]{derek.teaney@stonybrook.edu}
\affiliation{Department of Physics and Astronomy, Stony Brook University, Stony Brook, New York 11794, USA}
%\author{Fanglida Yan}
%\email[]{yan.fanglida@stonybrook.edu}
%\affiliation{Department of Physics and Astronomy, Stony Brook University, Stony Brook, New York 11794, USA}
\date{\today}
%%%%%%%%%%%%%%%%%%%%%%%%%%%%%%%%%%%%%%%%%%%%%%%%%%%%%%%%%%%%%%%%%%%%%%%%%%%

\begin{abstract}
We determine the kinetic coefficients of ultrasoft pions using chiral perturbation theory at finite temperature close to the chiral limit.  This is used to compute the axial charge diffusion and damping coefficients in the hydrodynamic effective theory for these pion waves. We show that to provide a leading order answer for these coefficients one needs to explore the dynamics of hard, soft, and ultrasoft pion modes, which are represented microscopically by the appropriate kinetic and hydrodynamic descriptions.
\end{abstract}

%\keywords{}

\maketitle

\clearpage

\section{Introduction}\label{intro}

The thermodynamic state of Quantum Chromodynamics (QCD) at low temperatures and zero quark mass is characterized by a broken chiral symmetry.  Specifically in the limit of two massless flavors the $O(4) \simeq SU_L(2) \times SU_{R}(2)$  symmetry is spontaneously broken to $SU_V(2)$. Since the associated Goldstone bosons (the pions) have arbitrarily long wavelength, they should be included into the hydrodynamic description of the microscopic theory. We denote these ``hydrodynamic pions'', as opposed to hard modes which, after coarse graining, become part of the conserved hydrodynamic fields. The resulting ideal equations of motion with the broken symmetry resemble a non-Abelian $O(4)$ superfluid~\cite{Son:1999pa}. 

At first order in the hydrodynamic gradient expansion, when the coupling between the pions and the conserved $O(4)$ currents is retained, but the coupling to the energy-momentum tensor $T^{\mu\nu}$ is neglected,  
the superfluid theory has two dissipative coefficients, which characterize the damping of the soft pion waves in the effective theory~\cite{Son:2001ff,Son:2002ci,Grossi:2020ezz}. 
Our immediate goal is mainly theoretical: we wish to compute these dissipative coefficients in a  microscopic theory, which realizes the broken chiral symmetry and has a calculable chiral limit.  For this purpose we will work with finite temperature chiral perturbation theory ($\chi$PT), with an arbitrarily small pion mass, and show how these dissipative coefficients can be computed. We hope that the concrete computation of the dissipative coefficients in this case will clarify the physical content of the $O(4)$ superfluid theory more generally. 

When the quark mass is small, but finite, the pion acquires a small mass $m$ with a Compton wavelength $\lambdabar \sim \hbar c/m$.  At very large distances $L \gg \lambdabar$, the appropriate effective theory is ordinary hydrodynamics, while at shorter distances $L \sim \lambdabar$, the pions need to be explicitly propagated as hydrodynamic degrees of freedom~\cite{forster2018hydrodynamic,Son:1999pa}.  For the pion hydrodynamic picture to be valid, the Compton wavelength $\lambdabar$ should be large compared to the typical mean free path of the system, $\lambdabar \gg  \ell_{\rm mfp}$. 
When the soft (or superfluid) degrees of freedom are further integrated to find an effective theory  $L\gg \lambdabar$, these soft pions give  finite and computable  contributions to the transport coefficients of the ordinary hydrodynamic theory. These contributions are expressed in terms of the dissipative parameters of the superfluid effective theory~\cite{Grossi:2020ezz}. In particular, for the ordinary isospin conductivity $\sigma_I$, this superfluid contribution is in fact the dominant one, and is inversely proportional to the pion mass~\cite{Grossi:2020ezz}.
 
Thus, by finding the dissipative parameters of the superfluid pion effective theory, we will, as a by-product, determine the isospin conductivity in chiral perturbation theory to leading order in the pion mass. Similarly, we will also be able to determine the leading contribution to the ordinary bulk viscosity  in a specific
temperature range, which is sensitive to the soft pion modes~\cite{Grossi:2020ezz} (as we will see, when arbitrarily soft pions are allowed the  effective $1 \rightarrow 3$  vertices become kinematically allowed,  and play a role in determining the bulk viscosity). For the shear viscosity, $\eta$, the superfluid theory determines the leading dependence of $\eta$ on the pion mass.

There exist several previous calculations of the kinetics of $\chi$PT. In most of the literature the connection with the underlying superfluid theory was poorly understood; these connections were made explicit later by Son~\cite{Son:1999pa} and Son and Stephanov~\cite{Son:2001ff,Son:2002ci}. In ideal superfluid hydrodynamics, the pion velocity at zero momentum $v_0^2(T)=f^2/\chi_A$ in the chiral limit plays an important role~\cite{Son:2002ci} ($f$ is the pion decay constant and $\chi_A$ the axial susceptibility). $v_0^2(T)$ has been calculated in an early insightful paper by Schenk~\cite{Schenk:1993ru}, and later elaborated on by Toublan~\cite{Toublan:1997rr}.

Previous computations of transport coefficients in $\chi$PT have tacitly assumed that the pion Compton wavelength is short compared to the mean free path, $\lambdabar \ll \ell_{\rm mfp} =T^{-1} (f/T)^4$~\cite{Gasser:1986vb,Gerber:1988tt,Dobado:2003wr,FernandezFraile:2005ka,FernandezFraile:2009mi,Lu:2011df,Dobado:2011qu,Torres-Rincon:2012sda}. In this regime  of thermal $\chi$PT the pion momentum, the pion mass\footnote{In this case the explicit chiral symmetry breaking term in the chiral Lagrangian is proportional to $m^2$, and is already included at leading order. }, and the temperature are the same order of magnitude
\st 
p,m \sim T  \, ,
\stp
and all of them need to be much smaller than the chiral scale,
\st
 \Lambda_\chi = 4\pi f \simeq 1 \textrm{ GeV} \ .
\stp
In this regime one naturally has
\st 
\frac{\lambdabar}{\ell_{\rm mfp}}= \frac{T^5}{m f^4} \ll 1 \ .  \label{eq:scale}
\stp

Indeed in this regime, classical kinetics with massive particles  is appropriate  and the transport computations are conceptually straightforward, either solving the Boltzmann-Uehling-Uhlenbeck equation or via the Green-Kubo approach.

In this work we will consider the small mass limit, requiring that the mass is smaller than the collision rate,  $\lambdabar \gg \ell_{\rm mfp}$\,  or \, $m \ll T (T/f)^4$. Taking the small mass limit does not invalidate $\chi$PT of
course, and one can still use the chiral Lagrangian in a  microscopic analysis of the kinetics~\cite{Gerber:1988tt}.  However, having $\lambdabar/\ell_{\rm
mfp} \gg 1$ leads to a rich interplay of scales; infrared divergences are not
simply cut off by the pion mass, and  an interesting resummation is required to
determine the damping rates of soft pions. This resummation was been
anticipated by Smilga, who estimated the damping rate  at leading log in
$T/f$~\cite{Smilga:1996cm}. However, a detailed analysis of the kinetics of the
soft pion fields (which is required to find the coefficient under the log)
has never been given.

The regime of $\chi$PT with $\lambdabar \gg \ell_{\rm mfp}$ is certainly quite far from the real world. However, at experimentally relevant  temperatures (with $T$ approaching the crossover temperature), the pion Compton wavelength in the Hadron Resonance Gas (HRG)  is becoming comparable to the spacing between hadrons, $\lambdabar \gtrsim \ell_{\rm mfp}$. In an idealization of this regime, the emergent physical picture consists of  hard hadron modes moving in a background of ultrasoft pion fields, which  are described by the dissipative superfluid dynamics.  Similarly, in $\chi$PT for $\lambdabar \gg {\ell}_{\rm mfp}$,  the hard pion modes with $p\sim T$ propagate in a background field of soft pions~\cite{PhysRevLett.78.3622,Manuel:1997zk}.  By calculating the dissipation rates of the soft pion fields within the $\chi$PT context we hope to provide guidance to the much more complicated HRG. Indeed, it should be possible to extend our results to the interacting HRG at higher temperatures by using experimental inputs on the cross sections of hadrons, and by enforcing chiral constraints on the interactions of soft pions and hard hadronic modes. In addition, the possibility to continuously take the chiral limit is of much interest for studying the $O(4)$ transition, as it has been done in recent lattice-QCD calculations~\cite{HotQCD:2019xnw,Kaczmarek:2020sif}, with a pion mass less than half of its physical value, where the condition \Eq{eq:scale} can be met.

While this paper is primarily of theoretical interest, let us briefly describe the phenomenological motivations for this work.  Collisions of heavy ions at the Relativistic Heavy Ion Collider (RHIC) and Large Hadron Collider (LHC) are remarkably well described by viscous hydrodynamics, which predicts the measured flow harmonics and their correlations in exquisite detail~\cite{Jeon:2015dfa,Heinz:2013th}.  Current hydrodynamic simulations in this regime are based on a  theory  of ordinary hydrodynamics, which ignores chiral  symmetry breaking at low temperatures and the associated superfluid dynamics.  This is reasonable,  since the quark mass is finite in the real world and  chiral symmetry is always explicitly broken.  Nevertheless, if the quark mass is small, one would expect that the pattern of chiral symmetry breaking could provide a useful organizing principle for hydrodynamics, increasing its predictive power. Indeed, in the small mass limit the system passes close to the $O(4)$ critical point of the system~\cite{Ding:2019prx}. The critical $O(4)$ dynamics then matches smoothly below $T_c$ with the superfluid theory described here.  Experimentally, it is found that the soft pion yields are enhanced relative to the predictions of ordinary hydrodynamics (see for example~\cite{Devetak:2019lsk}). It is hoped that critical $O(4)$ dynamics can describe both the enhanced pion yields, and the still-to-be-measured correlations among the soft pions.

An outline of the paper is as follows. First we review the superfluid theory of ultrasoft pions in \Sect{sec:eom}, in order to precisely define the dissipative coefficients which we wish to compute from the microscopic Lagrangian. Next in \Sect{sec:chiptoverview} we review finite temperature $\chi$PT close to the chiral limit. In \Sect{sec:hardpions} we review some results on pion velocity and damping rate for hard modes. In \Sect{sec:soft} we will also introduce the soft and ultrasoft (or hydrodynamic) scales, which will give structure to the subsequent computations in \Sect{sec:transportcoeffs}.
In \Sect{sec:softdetailed} we determine the contribution of the hard modes to the damping rate of  hydrodynamic pion waves.
Even the after including the quasiparticle width on the hard lines (which is essential to a correct treatment),  the hard contribution is infrared divergent, indicating a sensitivity to the soft sector. Thus in \Sect{sec:resummedsoft} the soft contribution to the damping rate is analyzed. The analysis is greatly simplified by using a sum rule familiar from high temperature QCD plasmas~\cite{Caron-Huot:2008zna,Ghiglieri:2013gia}. When the hard and soft contributions are combined in \Sect{sec:axialdiff} the final result for the damping rate is independent of the separation scale dividing these two descriptions.  Finally, a summary and outlook is presented  in \Sect{sec:outlook}.

\section{Hydrodynamic equations close to the chiral limit}
\label{sec:eom}

In this section we summarize the hydrodynamic equations of motion for a pion wave, in order to properly define the relevant kinetic coefficients. The two coefficients which will be computed in this work are the {\it axial charge diffusion coefficient} $D_A$, and the {\it axial damping coefficient} $D_m$. The pion wave propagates with an ultrasoft (or hydrodynamic) momentum, i.e.  its wavelength is long compared to the typical mean free path. We follow the notation and conventions of Ref.~\cite{Grossi:2020ezz}.
  
We consider the partially conserved axial current (PCAC) relation\footnote{Notice that $J_A^\mu=(J_A)^\mu_a t^a$ where $t^a$ are the generators of the $SU(2)$ algebra, normalized to $\textrm{Tr } [t^at^b]={\cal T}_F \delta^{ab}$, with ${\cal T}_F=1/2$. The current in question is the isoaxial current $(J_A)^\mu_a = \langle \bar{\psi} \gamma^{\mu} \gamma^5 t_a \psi \rangle$.},
\st \partial_t J_A^0 + \nabla \cdot {\bm J}_A = f^2 m^2 \varphi \  , \label{eq:PCAC} 
\stp
where $f$ and $m$ are the pion decay constant and screening mass (we reserve $F$ and $M$ for the parameters of the $\chi$PT Lagrangian). At ideal order the components of the axial current are related to the pion field $\varphi$ as
\begin{align}
J_A^0 & =\chi_A \mu_A \ , \\
{\bm J}_A & = f^2 \nabla \varphi \ , \label{eq:axialcurrent}
\end{align}
where $\chi_A$ is the axial charge susceptibility, and the relation between $\mu_A$ and $\varphi$ is given by the Josephson constraint, which in the ideal case reads $-\pa_t \varphi =\mu_A$. 

We assume that the PCAC relation also holds in the dissipative case~\cite{Son:2002ci,Grossi:2020ezz}, which amounts to a choice of a hydrodynamic frame\footnote{ This choice was made by Son and Stephanov whom we follow~\cite{Son:2002ci}. The operator equation in QCD is 
$ \partial_\mu (J_A)^{\mu,a}   
=  2 i m_q \bar \psi \gamma^5 t^a  \psi $.  Inserting the equilibrium chiral condensate $\llangle \bar \psi \psi \rrangle$, and using the equilibrium Gell-Mann--Oakes--Renner relation $f^2 m^2=-m_q \llangle\bar \psi \psi \rrangle$, we  define the field
  $\varphi^a \equiv  -2 i \bar\psi \gamma^5 t^a \psi/\llangle \bar \psi \psi \rrangle $. With this definition there are no dissipative corrections to the right hand side of the PCAC relation.
}. Then, we accommodate the effects of dissipation into the Josephson relation and the constitutive relation for the  current
\begin{align} 
   \partial_t \varphi & = -\mu_A  + \zeta^{(2)} f^2 \left(   \nabla^2 \varphi-  m^2 \varphi \right)  \ , \label{eq:dtphi} \\  
{\bm J}_A  & = f^2 \nabla\varphi  - \sigma_{A} \nabla\mu_A \label{eq:JA} 
\ , 
\end{align}
where the two transport coefficients ($\zeta^{(2)}$ and $\sigma_A$) are introduced upon the expansion in powers of gradient and pion mass~\footnote{While in Ref.~\cite{Son:2002ci} two independent dissipative coefficients were introduced in the Josephson relation (\ref{eq:dtphi}) in front of $\nabla^2$ and $m^2$, in~\cite{Grossi:2020ezz} it is shown that the positive entropy production constraint requires them to be equal.
See also the recent works~\cite{Delacretaz:2021qqu,Armas:2021vku} and references therein for discussions on this result in other systems. The latest authors find an additional parameter characterizing the residue of the pion pole. Here we are focused on the  pole position which is determined by the two kinetic coefficients described here.}.
Inserting this equation into the PCAC relation (\ref{eq:PCAC}) we get
\st \chi_A \partial^2_t \varphi  -f^2 \nabla^2 \varphi +f^2m^2\varphi - \lambda_A \nabla^2 \partial_t \varphi + \lambda_m m^2 \partial_t \varphi = 0 \ , \label{eq:wave}
\stp
where we have defined $\lambda_A \equiv D_A \chi_A$ and $\lambda_m \equiv D_m \chi_A$, with~\cite{Grossi:2020ezz}
\begin{align}
   \label{eq:sigmarelations}
  D_A & \equiv \sigma_A/\chi_A+f^2 \zeta^{(2)} \ , \\
  D_m & \equiv f^2 \zeta^{(2)} \ .
\end{align}
Equation~(\ref{eq:wave}) can be interpreted as the equation of motion for ultrasoft pions in the presence of dissipation. Clearly the coefficients $\lambda_A,\lambda_m$ cause a damping of the propagating pion wave. Assuming the latter in the following form $\varphi \sim \exp(-i\omega t+i {\bm q} \cdot {\bm x})$ we find the following dispersion relation,
\st  -\omega^2  + v^2_0 ( \q^2 + m^2) - i \omega \Gamma_q  = 0 \ , \label{eq:disprel}
\stp
where we have defined $\Gamma_q \equiv D_A \q^2 +D_m m^2$, and introduced the pion velocity $v^2_0=f^2/\chi_A$~\cite{Grossi:2020ezz}. Notice that for terms to be of the same order we have $m, \omega \sim {\cal O} (\q)$ and $v_0 \sim {\cal O}(1)$. In particular, the mass is considered to be an ultrasoft scale (see the general discussion in Sec.~\ref{intro}).

The dissipative coefficients are computed through the medium modifications of the retarded pion propagator, which should match with the hydrodynamic prediction
\st G^R (\omega,{\bm q}) = \frac{1}{\chi_A} \frac{1}{-\omega^2 +\omega_q^2 -i\omega \Gamma_q} \ , 
\stp
where $\omega_q^2 \equiv v^2_0 (\q^2+m^2)$. After matching to the generic form including the pion self-energy,
\st G^R (\omega,{\bm q}) \propto \frac{1}{-\omega^2 + \omega^2_q - \Sigma^R(\omega,{\bm q})} \ , 
\stp
one can identify in the small-${\bm q}$ limit
\st \Gamma_q = \frac{1}{\omega} \textrm{Im } \Sigma^R(\omega,{\bm q}) > 0 \ , 
\stp
where the external $\omega$ is evaluated on shell,  $\omega =\omega_q$. We have chosen to extract the parameters of the hydrodynamic theory from the correlations of the Goldstone fields, but we could have equally extracted them from the correlations of axial current. Indeed these quantities are related by Eq.~(\ref{eq:axialcurrent})  and this leads to simple relations between the correlators of the axial charge and the Goldstone fields (see Eqs.~(70) and (71) in Ref.~\cite{Grossi:2021gqi}.)

Therefore the kinetic coefficients $D_A$ and $D_m$ can be computed from the imaginary part of the self-energy when the external momentum is ultrasoft. In particular, the axial charge diffusion can be computed in the chiral limit  limit $m=0$,
\st D_A = \lim_{\q \rightarrow 0} \left. \frac{1}{\q^2 \omega}  \textrm{Im } \Sigma^R (\omega,{\bm q})  \right|_{\omega = v_0 q}\ . \label{eq:DA} 
\stp
Taking the mass finite, but ultrasoft, the coefficient $D_m$ can be determined from the self-energy as
\st D_m = \lim_{\q \rightarrow 0}  \left. \frac{1}{m^2\omega} \textrm{Im } \Sigma^R (\omega, {\bm q}) \right|_{\omega = v_0 m}  \ .
\label{eq:Dm} 
\stp

The calculation of $D_A$ and $D_m$ is performed in \Sect{sec:transportcoeffs}. In practice, we will compute the 
Wightman self-energy $\Sigma^>(Q)$, which is related to the imaginary part of the retarded self-energy through the fluctuation-dissipation theorem (FDT),
\st
\frac{ \textrm{Im} \Sigma^R(\omega, {\bm q}) }{\omega} = \frac{\Sigma^>(\omega, {\bm q})}{2T} \, ,
\stp
and has a convenient set of cutting rules~\cite{Caron-Huot:2007zhp}.

The pion velocity $v^2_0$ is determined from the real part of the pion self-energy and has been computed previously as described in the next section~\cite{Schenk:1993ru,Toublan:1997rr}. There we  will review $\chi$PT at finite temperature more generally, and define the hard, soft, and ultrasoft scales more precisely.

\section{ $\chi$PT at finite temperature and separation of scales}
\label{sec:chiptoverview}

\subsection{Scattering and dispersion relation of hard pions: a conflict with hydrodynamics}\label{sec:hardpions} 

According to the previous section our main task is to evaluate the imaginary part of the pion self-energy at asymptotically small momentum ${\bm q}$. Before doing that, in this section we will evaluate the pion self-energy for hard on shell external momentum, $P=(p^0,{\bm p})$ with $p^0=|{\bm p}|\sim T$. This will set the notation and allow us to extract some quantities which will be needed for the main calculation. We will also see that the external momentum $p$ cannot be naively extrapolated to zero to find the transport coefficients in the ultrasoft limit.

The microscopic Lagrangian of massless $\chi$PT at next-to-leading order (NLO) is~\cite{Gasser:1983yg}
\begin{multline}
   \label{eq:chilagrangian}
\mathcal{L} = - \frac{F^2}{4} \Tr\left[ \partial_{\mu} U \partial^{\mu} U^\dagger \right] 
+ L_1 ( \Tr\left[ \partial_{\mu} U \partial^{\mu} U^\dagger \right])^2 
+ L_2 \Tr[\partial_{\mu} U \partial_{\nu} U^\dagger] \, \Tr[\partial^\mu U \partial^\nu U^\dagger]  \, .
\end{multline}
The low-energy coefficients (LECs) $L_1$ and $L_2$ will be unimportant 
for the computation of the dissipation rates, but they do affect the velocity $v_0$, which deviates from unity only at NLO~\cite{Schenk:1993ru,Toublan:1997rr}. As emphasized in 
the introduction, we are working asymptotically close to the chiral limit, so the mass term in the chiral Lagrangian has been neglected for the moment. 
The parameter in the Lagrangian $F$ is equal to the spatial decay constant $f$ 
introduced in the previous section at lowest order, but deviates at higher orders~\cite{Toublan:1997rr}.

\begin{figure}[ht]
   \begin{center}
 \includegraphics[width=0.36\textwidth]{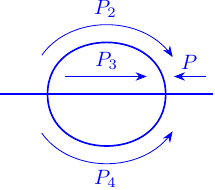}
 \end{center}
 \caption{\label{fig:hardself} Two-loop self-energy of hard pions (denoted in solid blue line). This gives the main contribution to the scattering rate (imaginary part), and a correction to the pion dispersion relation (real part).}
\end{figure}

A typical pion at finite temperature has momentum $p\sim T$ 
and we will refer to this momentum scale as hard. The relevant properties of these hard quasiparticles can be found by evaluating the retarded self-energy, which determines the scattering rate and the dispersion curve. 
The scattering rate of an on-shell pion with momentum $p$ is defined as 
\st
\gamma_p \equiv \frac{{\rm Im} \ \Sigma^R(p)}{p} \, ,
\stp
and the leading contribution to the imaginary part 
is given by the two-loop graph of \Fig{fig:hardself}~\cite{Goity:1989gs,Schenk:1993ru}, where here and below the solid blue lines  denote hard pions. We give details of this and other computations in Appendix~\ref{app:hard}, and here just summarize the relevant features of the result. 

\begin{figure}[ht]
   \begin{center}
  \includegraphics[scale=0.65]{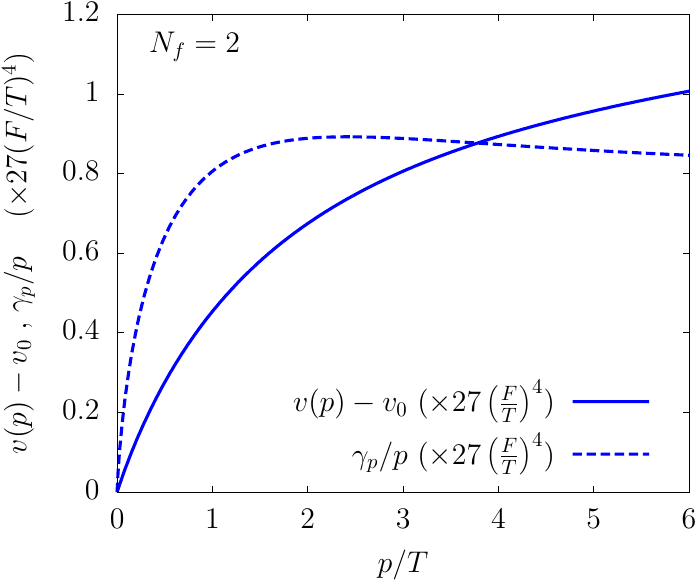}
 \end{center}
 \caption{\label{fig:quasiparticles} 
 Quasiparticle properties, velocity and damping rate, for hard pion modes of order $T$ in $SU(2)$ $\chi$PT. }
\end{figure}

The scattering rate is determined by the squared Weinberg amplitude and is of order 
\st
\gamma_p \sim  T \left( \frac{T}{F} \right)^4 \ .
\stp
A straightforward numerical evaluation of $\gamma_p$ is shown in  \Fig{fig:quasiparticles} by the  dashed line.  For later use we define 
the average scattering rate~\cite{Goity:1989gs,Smilga:1996cm},
\st
\avg = \frac{\int \frac{d^3p}{(2\pi)^3} n_p\gamma_p }{\int \frac{d^3p}{(2\pi)^3} n_p} = 2.33  \left(\frac{T^5}{27 F^4}\right) \ , \label{eq:avthermalgamma}
\stp
where $n_p$ is the Bose-Einstein distribution function, and the coefficient $2.33$ is the outcome of the numerical calculation.

At small momentum the scattering rate of pions is reduced due to their Goldstone character, and evaluating the imaginary part for $p\ll T$ (but still hard) we find
\st
\label{eq:gammasoftbad}
\lim_{p/T \rightarrow 0} \gamma_p =   \frac{p^2 T^3}{9\pi F^4} \log\left(1.56 \frac{T}{p} \right) \ ,
\stp
where the coefficient $1.56$ is determined numerically.
As emphasized by Smilga~\cite{Smilga:1996cm}, the logarithm in \Eq{eq:gammasoftbad} is inconsistent with the hydrodynamic 
expectation which predicts that $\gamma_p \propto p^2$. In the hydrodynamic limit, when $p$ becomes ultrasoft, the scattering rate should turn into $\Gamma_q$ defined in Eq.~(\ref{eq:disprel}). Indeed, a naive application of Eq.~\eqref{eq:DA} would lead to a logarithmically momentum dependent diffusion coefficient $D_A$, instead of a constant. As anticipated by Smilga---but not fully calculated---the solution to this problem is to cut off the momentum in the logarithm when it becomes comparable to the scattering rate.  This leads to an estimate for $D_A$ of order~\cite{Smilga:1996cm}
\st
D_A \sim  \frac{T^3}{F^4} \log \left( \frac{T}{\avg} \right) \ . 
\stp 
The remaining sections will discuss how to regulate the logarithm and how to determine the constant under the log consistently, as well as addressing  $D_m$ in a similar manner.  

Now we turn to the real part of the self-energy which determines the velocity of the hard pions in the medium. The full evaluation of this self-energy follows Schenk~\cite{Schenk:1993ru} and is discussed in Appendix~\ref{app:hard}. The dispersion 
curve of the hard pions can be written 
\st
   E_p  = v(p) p \ .
\stp
At small momentum the Goldstone character of the modes dictates that the dispersion curve should (in the chiral limit) take the form
\st
   E_p = v_0 p \ ,
\stp
where $v_0^2\equiv f^2/\chi_A$ is a constant that is determined by the equilibrium properties of the medium~\cite{Son:2002ci}. In contrast to the imaginary part of self-energy,  there is no obstacle in extrapolating the real part to zero momentum, and the resulting limiting velocity  $v_0$,  will be  recorded in the next paragraph. 
The difference $v(p) - v_0$
is independent of the low energy constants $L_1$ and $L_2$, and results solely from the
two-loop graph shown in Fig.~\ref{fig:hardself}. Indeed, dispersion relations relate $v(p) - v_0$  to the imaginary part of the self-energy which also is determined only by Fig.~\ref{fig:hardself} at leading order. The solid line in Fig.~\ref{fig:quasiparticles} records $v(p) - v_0$ which is the same order of magnitude as the collisional width.   

Both  $\gamma_p$ and the dispersion $v(p) - v_0$ will be important in the computations below.
Although we will not need the limiting velocity $v_0$ here, we  will record its value, since it provides a good 
check of our numerical work and is of considerable interest. Schenk~\cite{Schenk:1993ru} and Toublan~\cite{Toublan:1997rr} evaluated $v_0$ with different methods and found
\st
v_0^2 = 1 - \frac{T^4}{27 F^4} \log \left( \frac{\Lambda_{\Delta}}{T} \right) \ ,
\stp
where  $\log \Lambda_{\Delta}$  is a specific combination of LECs
\st
\log \left(\frac{\Lambda_{\Delta}}{\mu} \right)  = \frac{192\pi^2}{5} \left[ L_1^r(\mu) + 2 L_2^r(\mu) \right] + 0.54 \ , \label{eq:loglambda}
\stp
and $L_1^r(\mu)$ and $L_2^r(\mu)$ are the conventional dimensionally 
regularized LECs~\cite{Gasser:1983yg,Toublan:1997rr}, 
\st
\label{eq:Lis}
L_i = L_{i}^r(\mu)  + \gamma_i \lambda \, ,
\stp
with  $\gamma_1 = 1/12$ , $\gamma_2 = 1/6$, and 
\st
\lambda = \frac{\mu^{d-4}}{32 \pi^2} \left( \frac{2}{d - 4} - \log(4\pi) + \gamma_E - 1\right) \ .
\stp
The constant $0.54$ in Eq.~(\ref{eq:loglambda}) is consistent with our numerical work.
Many years ago Toublan made a rough estimate of $\Lambda_\Delta \simeq 1.8\,{\rm GeV}$
by using the measured low energy constants $L_1^r$ and $L_2^r$. He then estimated
the chiral phase transition temperature as $T \simeq 160\, {\rm MeV}$ 
by extrapolating $v_0$ to zero~\cite{Toublan:1997rr}. This estimate is roughly consistent with current lattice-QCD measurements of the chiral crossover temperature of $T_\chi=151(3)(3)$ MeV~\cite{Aoki:2006br}.

\subsection{Hard, soft, and ultrasoft: the need for a resummation~\label{sec:soft}}

\begin{figure}[ht]
  \centering
  \includegraphics[width=65mm]{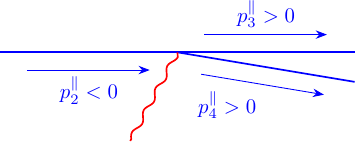}
  \caption{Splitting of a hard pion into two almost collinear pions, when one of the incoming pions (in red wavy line) is ultrasoft. }
  \label{fig:split}
\end{figure}

Consider the binary scattering of two pions with incoming momenta $Q$ and
$P_2$, and outgoing $P_3$ and $P_4$. Assume that all momenta are hard ($\sim
T$) except for $Q$ which is taken as an ultrasoft (or hydrodynamic) pion. The
meaning of ultrasoft will be defined below. The collision is depicted in
Figure~\ref{fig:split} (together with the sign conventions), and the ultrasoft
pion is represented by a red wavy line. As $Q \ll P_2,P_3,P_4$ this diagram
corresponds to an effective $1\rightarrow 2$ splitting process, and the three
hard pions are nearly collinear. The momentum along the collinear axis is notated
$p^\pl_i$, with a negative $p^\pl_i$ indicating a particle in the initial state
such that $p_2^\pl + p_3^\pl + p_4^\pl = 0$.

\begin{figure}[ht]
  \centering
  \begin{minipage}[c]{0.33\textwidth}
  \includegraphics[width=\textwidth]{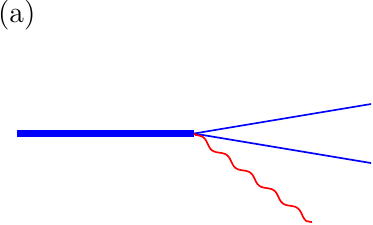}
  \end{minipage}
  \hspace{0.08\textwidth}
  \begin{minipage}[c]{0.47\textwidth}
  \includegraphics[width=\textwidth]{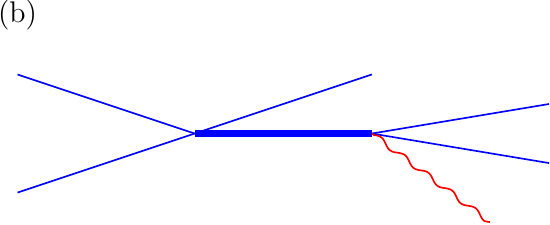}
  \end{minipage}
  \caption{(a) An approximately on-shell pion decaying into two nearly on-shell collinear pions and an ultrasoft mode.  (b) An example of a $2\rightarrow 4$ scattering process 
  captured  by (a).}
  \label{fig:twotothree}
\end{figure}

The energy difference of the process gives an estimate of the inverse collision duration,
 \st 
 \label{eq:deltaE22}
 \delta E = E_3 +E_4 - E_Q - E_2 \simeq (v_3-v_2) p_3^\pl + (v_4-v_2) p_4^\pl \ , \stp 
where we have used the dispersion relation at lowest order,  $E_i \simeq v_i p_i^\pl$ with $v_i = v(|p_i^\pl|)$.
%with an obvious notation $v_i \equiv v(p_i^\pl)$. We have neglected 
The ultrasoft momentum $Q$ has been dropped, since (as discussed further below) it is parametrically small compared to the difference  $\delta E = E_3 + E_4  - E_2$.
Numerical results for the velocity differences, $v(p) - v_0$, were presented 
in  \Fig{fig:quasiparticles}, and are of order
 \st 
 v(p_i) - v_0 = {\cal O} \left( \frac{T^4}{F^4} \right) \ . 
\stp
We see that the ``off-shellness'' (or inverse collision duration) of the splitting process is of order 
\st \delta E \sim T \frac{T^4}{F^4}  \, ,  \stp 
which is the same order as the collisional rate of the hard pions, cf. Eq.~(\ref{eq:avthermalgamma}). Thus, the hard pions involved in the process need to be dressed by their on-shell self-energies. 
Hard pion propagators, with the collisional width incorporated, will be diagrammed as a wider blue line as in \Fig{fig:twotothree}. 

We can now clarify our terminology for hard, soft, and ultrasoft momenta, which we typically denote with $P$, $K$, and $Q$ respectively. 
Pions with momentum $P \sim T$ are called hard, those with momentum of order the collisional width  $K \sim T (T/F)^4$  are called soft, and those with  momentum  $Q \ll T(T/F)^4$, are called ultrasoft or hydrodynamic. 

Because the ultrasoft momentum 
played no role in the energy budget described by \Eq{eq:deltaE22}, the ultrasoft pion
can be either in the initial state, as depicted in \Fig{fig:split}, or in the final state, as depicted in \Fig{fig:twotothree}~(a). Physically this graph represents an almost on-shell pion decaying into two hard collinear pions, while increasing the amplitude of a background ultrasoft wave.  More concretely \Fig{fig:twotothree}~(a) represents, for example,  a hard $2\leftrightarrow 2$ collision in the past producing an almost on-shell quasiparticle, which ultimately splits into two collinear pions, as depicted in \Fig{fig:twotothree}~(b). 

In the next section we will analyze the self-energy of ultrasoft pions
in detail, and adhere to  this discussion by including  the collisional width of the hard lines.

\section{The pion self-energy and the kinetic coefficients}
\label{sec:transportcoeffs}

Having completed some preliminary remarks about chiral perturbation
theory at finite temperature in \Sect{sec:chiptoverview}, in this section we will evaluate the pion self-energy when the external momentum is hydrodynamic, and determine the kinetic coefficients $D_A$ and $D_m$.
In \Sect{sec:softdetailed}, we will first evaluate this self-energy when 
all of the internal lines are hard, $P\sim T$, including the collisional width as explained in \Sect{sec:soft}. The result is infrared divergent, indicating a sensitivity to the soft sector,  when one of internal lines becomes soft, $K \sim T(T/F)^4$. In \Sect{sec:resummedsoft} we compute the self-energy in this soft kinematic regime, exploiting a sum-rule technique familiar from QCD plasmas at high temperature~\cite{Caron-Huot:2008zna,Ghiglieri:2013gia}. The sum of the two partial results 
for the self-energy is presented in \Sect{sec:axialdiff}, and is independent of the cutoff separating the two scales. This result realizes the resummation anticipated by Smilga~\cite{Smilga:1996cm}.

\subsection{The resummed self-energy: contribution from hard modes~\label{sec:softdetailed}}

\begin{figure}
  \centering
  \includegraphics[width=68mm]{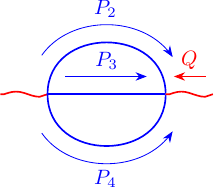}
  \caption{Two-loop self-energy of an ultrasoft pion. The internal lines are
  hard and dressed by their on-shell self energies.}
  \label{2loop_usoft}
\end{figure}

We start with the self-energy diagram of Fig.~\ref{2loop_usoft}, where all internal momenta are hard,  and all momenta are flowing into the first vertex.  As discussed in \Sect{sec:softdetailed}, this is effectively a $1 \rightarrow 2$ splitting process and  its inverse (a $2 \rightarrow 1$ joining process) in an ultrasoft pion background.  The splitting and joining rates depend on $\delta E/\gamma$, where $\gamma$ is the collisional width of the hard lines. Our goal in this section is to derive \Eq{eq:splitjoin} which makes this interpretation explicit.

The self-energy $\Sigma_{ab}^>(Q)$ reads 
\begin{multline} 
   \Sigma_{ab}^>(Q)  =   \frac{1}{2!} \int \frac{d^4P_2}{(2\pi)^4}
\frac{d^4P_3}{(2\pi)^4} \frac{d^4P_4}{(2\pi)^4} (2\pi)^4 \delta^{(4)} (Q+P_2+P_3+P_4) \sum_{a_2,a_3,a_4} i{\cal M}_{aa_2,a_3a_4} (i{\cal M}_{ba_2,a_3a_4} )^*  \\
\times G^>(P_2)G^>(P_3)G^>(P_4) \label{eq:pionselfenergy}
\end{multline}
where $a,b$ are isospin indices, and the invariant scattering amplitude for the usual $SU(2)$ case is
\st i {\cal M}_{a_1a_2a_3a_4} = \frac{i}{F^2} \left[ \delta_{a_1a_2} \delta_{a_3 a_4} (-2 Q \cdot P_2)  + \delta_{a_1 a_3} \delta_{a_2 a_4} \left(-2 Q \cdot P_3 \right)  + \delta_{a_1 a_4} \delta_{a_2 a_3} \left(-2 Q \cdot P_4 \right) \right] \ . \label{eq:invampl} \stp

In terms of the average scattering amplitude squared (see App.~\ref{app:hard} for details), we have
\st \sum_{a_2,a_3,a_4} i{\cal M}_{aa_2,a_3a_4} (i{\cal M}_{ba_2,a_3a_4} )^* =  \delta_{ab} \overline{|\mathcal M|^2} \ , \stp
with
\st
\overline{|\mathcal M|^2} = \frac{1}{d_A} \sum_{a_1a_2a_3a_4} | \mathcal M^{a_1a_2}_{a_3a_4}|^2 
= \frac{2}{F^4} \left(s^2 +  t^2 + u^2 \right) \ , \label{eq:Maverage}
\stp
where for $d_A = 3$ (number of pions species) for two flavors. The Mandelstam variables are $s=(Q + P_2)^2$, $t=(P_3 + Q)^2$ and $u=(P_4 + Q)^2$.  

We note in passing that the Weinberg amplitude for massive pions in the general $SU(N)$ case reads (see details in App.~\ref{app:sun})
\st
\overline{|{\cal M}|^2}= \frac{1}{N^2-1} \sum_{a_1a_2a_3a_4} | \mathcal M^{a_1a_2}_{a_3a_4}|^2  = N^2 \frac{s^2+t^2+u^2}{2F^4} -2 \frac{N^4+2N^2-6}{N^2} \frac{m^4}{F^4} \ , \label{eq:M2}
\stp
where the dependence on the mass arises from the $-F^2 m^2 {\rm Tr} (U + U^\dagger) /4$ portion of the chiral Lagrangian. Equation (\ref{eq:M2})
reduces to the well-known result for $N=2$~\cite{Weinberg:1966kf,Gasser:1983yg}
\st
\overline{|{\cal M}|^2}= \frac{ 2 (s^2+t^2+u^2)-9m^4}{F^4} \ . \label{eq:M2SU2}
\stp

  It is important to note that the explicit mass term of Eq.~(\ref{eq:M2}) is of order $m^4$, while the remaining terms are lower order, e.g.  $s^2 \sim Q^2 T^2$.
Here and below we will leave $Q=(\omega, {\bm q})$ arbitrary but ultrasoft, and only at the end will we set $\omega^2(q) = v_0^2 (\q^2 + m^2)$.  Indeed, the leading mass dependence of the damping rate comes from the
frequency dependence of $\Sigma^>(\omega,{\bm q}) = c_1 \omega^2 + c_2 \q^2$ and the on-shell dispersion relation for $\omega(q)$. All other mass corrections (e.g. on internal lines and vertices) are necessarily higher order.

Then, we define $\Sigma^>_{ab} (Q) = \delta_{ab} \Sigma^>(Q)$ and compute below
\begin{multline}
\Sigma^>(Q) = \frac{N^2}{4F^4} \int \frac{d^4P_2}{(2\pi)^4}
\frac{d^4P_3}{(2\pi)^4} \frac{d^4P_4}{(2\pi)^4} (2\pi)^4 \delta^{(4)} (Q+P_2+P_3+P_4) (-2Q\cdot P_2)^2 \\
\times G^>(P_2)G^>(P_3)G^>(P_4) \ , \label{eq:sigmaaux}
\end{multline}
for the $SU(N)$ case.

% The main contribution to the self-energy comes from hard internal pion lines, $Q$ being negligible.
%Softening of internal lines, makes the integral (\ref{eq:sigmaaux}) to diverge in the IR~\cite{Smilga:1996cm}. To regularize it, we introduce damping coefficients in the spectral function of hard pions $\gamma_p$, exploiting the quasiparticle picture described in Sec.~\ref{sec:hardpions}. 
The Wightman function for each of the hard pions reads,
\st G^>(P)= [ 1 + n(p^0)] \rho(P) = [1+n(p^0)] \ \sum_{s=\pm}  \frac{1}{2 E_p} \left( \frac{s \gamma_p}{ (p^0 - s E_p)^2 + (\gamma_{p}/2)^2 } \right) \ , \label{eq:Gpi} \stp
where $\rho(P)$ denotes the spectral function, and the correction to the dispersion relation
\begin{align}
 E_p\simeq &  v(p) |p_\pl| \left( 1 + \frac{p_\perp^2}{2 p_\pl^2 } \right) \ , \label{eq:harddisp}
\end{align}
is obtained in Sec.~\ref{sec:hardpions} by computing the two-loop modification of the real part of the hard pion self-energy.

After performing some integrations and reducing the integrand as indicated in App.~\ref{app:soft} one arrives to a compact result which precisely encodes the $1\leftrightarrow2$ splitting rates in the ultrasoft background field,
\begin{multline}
 \Sigma^>(Q)  =  \frac{N^2}{16\pi F^4} \left(\omega^2 + \tfrac{1}{3}\q^2\right) \int_{-\infty}^{+\infty} \frac{dp^\pl_3}{2\pi}  \int_{-\infty}^{+\infty} \frac{dp^\pl_4}{2\pi}   \int_{-\infty}^{+\infty} \frac{dp^\pl_2}{2\pi}   p^{\pl,2}_2 \ 2\pi \delta(p^\pl_2+p^\pl_3+p^\pl_4)  \\
 \times  (1 + n_2)    (1 + n_3)    (1 + n_4) s_{234}
      \ 2  \left[\frac{1}{2} +  \frac{1}{\pi} \tan^{-1} \left(  \frac{2\delta E}{\gamma} \right) \right] \ .  \label{eq:splitjoin}
\end{multline}
Here the $p^\pl$ are components of the collinear momenta of the $1\leftrightarrow 2$ splitting and joining processes along their common axis. A negative $p^\pl$ indicates particle in the initial state. $n_i \equiv n(p^\pl_i)$ are the Bose-Einstein functions, with $n(-p)=-1- n(p)$. The sign, $s_{234}$, is plus for joining processes, and minus for splitting processes, and balances the signs from the Bose-Einstein functions to make a positive integrand. Thus  for negative $p_2^\pl$ and positive $p_3^\pl$ and $p_4^\pl$,  we have
a splitting process with population factors
\st
  s_{234} (1 + n_2) (1 + n_3) (1 + n_4) = n_2 (1 + n_3) (1 + n_4) \, .
\stp
$\delta E$ is the energy difference (or inverse formation time) discussed above (see also Eq.~(\ref{eq:deltaE})) and $\gamma = \gamma_2+\gamma_3+\gamma_4$ is sum of the collisional widths.  
As anticipated in \Sect{sec:softdetailed} the splitting and joining rates in the soft background depend on the ratio of $\delta E$ to the collisional width.

The  energy difference $\delta E$ is related to the dispersion $v(p) - v_0$ by \Eq{eq:deltaE22}  and must be evaluated numerically, as well as the collisional width $\gamma(p)$.  Using
the numerical results for these quantities presented in \Fig{fig:quasiparticles}, the remaining numerical integrations are straightforward, and yield after \App{app:soft},
\st \Sigma^>(Q) = \frac{N^2 T^4}{24\pi F^4} (\omega^2 + \tfrac{1}{3} \q^2) \left[  \log \left( \frac{T}{\Lambda} \right) + 0.37 \right] \ , \qquad \textrm{(hard)} \label{eq:Sigmahard} \stp
The coefficient in front of the logarithm coincides with the result quoted in Ref.~\cite{Smilga:1996cm}. 
Examining this result we see that the damping rate of 
ultrasoft pions by hard  splitting and joining processes is logarithmically  
sensitive to an infrared cutoff $\Lambda$, which  excises  the $p_{3}^{\pl}$ and $p_4^{\pl}$ integrations in the IR. We will address this soft kinematic region in the next section.

Finally, we note that apart from the global $N^2$ factor of \Eq{eq:Sigmahard} the calculation of the coefficient under the logarithm makes an implicit use of the $SU(2)$ case, through the quantities appearing in Fig.~\ref{fig:quasiparticles}. Therefore a subleading $N$-dependence has been neglected in the result of \Eq{eq:Sigmahard}. However a full complete calculation for arbitrary $SU(N)$ is beyond the scope of this work.

\subsection{The resummed self-energy: a sum rule for soft modes}
\label{sec:resummedsoft}

In the previous section we found that the dissipation of ultrasoft pion waves by hard modes is logarithmically sensitive 
 to an integration region when one of the internal pion lines
 becomes soft. 
  
\begin{figure}[ht]
  \centering
  \includegraphics[width=140mm]{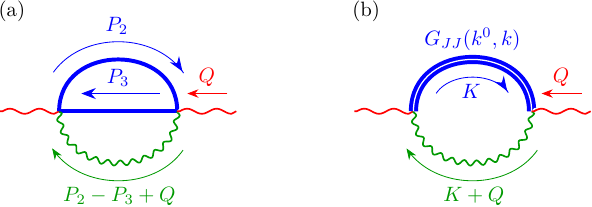}
  \caption{(a) The ultrasoft pion self-energy $\Sigma^>(Q)$ when the internal momentum $K\equiv P_2 - P_3$ and the green line are becoming soft.  (b) In the overlap region, $T(T/F)^4 \ll K \ll T$, the contribution of two hard lines in (a) can be recognized as the isovector current-current correlation function (double blue line).}
  \label{fig:2loopcorrel}
\end{figure}

Examine \Fig{fig:2loopcorrel}(a) and consider $K\equiv P_2 - P_3$  to be much smaller than $T$,  but still  large compared to $T(T/F)^4$, i.e. at the boundary of applicability of the hard analysis of the previous section.
We need to analyze the kinematics of the 2-loop diagram shown in the panel (a) of Fig.~\ref{fig:2loopcorrel}, where the hard pion lines already incorporate the collisional width, but one internal propagator is becoming soft (represented by the wavy green line). Because the external momentum is ultrasoft, the combined momenta of the two hard lines should be soft, to compensate the remaining soft line ($P_3$ should be almost opposed to $P_2$). In this limit it is convenient to recognize that the pair of hard lines is the isovector current-current correlator evaluated at the soft momentum $K$, as shown in panel (b) of Fig.~\ref{fig:2loopcorrel}, where the double solid blue line represents this correlator.  
For $K$ large compared to the scattering rate, but small compared to $T$, 
this current-current correlator can 
be evaluated using free kinetic theory, which is equivalent to a diagrammatic evaluation of the hard blue loop in panel (a).
The result is given in Appendix~\ref{app:correlator}.

Now consider $K$ to be fully soft $\sim T(T/F)^4$. The right panel of Fig.~\ref{fig:2loopcorrel} describes the soft contribution to the ultrasoft damping rate, and can be evaluated provided the isovector current-current correlator is known. This isovector correlator can be evaluated from straightforward kinetic theory, but this  evaluation should now incorporate collisions, since $K$ is of order the typical scattering rate.  As we will see, however, this evaluation is unnecessary due to a marvelous light-cone sum rule familiar from high temperature QCD plasmas~\cite{Caron-Huot:2008zna,Ghiglieri:2013gia}. Indeed, the graph in Fig.~\ref{fig:2loopcorrel} evaluates to an integral over the 
current-current correlator on the light cone, which in turn 
is the isospin susceptibility, up to a contact term that precisely matches with the logarithm of the previous section. The next subsections make this explicit.  The final results of these steps are presented in \Eq{eq:Sigmasoft}.

\subsubsection{Preliminaries}

We first express the (b) diagram of Fig.~\ref{fig:2loopcorrel} as an
integral over the isovector current-current correlator. 
The result is presented in \Eq{responseintegralform}, where $G_\longi$ and 
$G_\trans$ are the longitudinal and transverse components of the isovector
current-current correlator.

The isovector current reads, from the effective chiral Lagrangian,
\st J^{\mu,a} (X) = (T^a)^b_c \ \pa_\mu \pi_b (X) \pi_c(X) \ , \stp
where $(T^a)^b_c=f^{abc}$, and the symmetric  correlation function is
\st G_{JJ, \textrm{sym}}^{\mu\nu} (X,Y) = \frac12 \langle \{ J^\mu(X), J^\nu (Y)\} \rangle \ . \stp

The single diagram in panel (b) of Fig.~\ref{fig:2loopcorrel} reads
\st \Sigma^>_{ab} (Q)= \int \frac{d^4K}{(2\pi)^4} \left[ -i\frac{(T_a)^c_d}{2F^2} (2Q+K)_\mu \right] \left[ i\frac{(T_b)^d_c}{2F^2} (2Q+K)_\nu \right]  G_{JJ}^{\mu\nu,>} (K) G^> (Q+K)  \ .  \stp
We can use the FDT relations  at small momenta
\begin{align}
 G_{JJ}^{\mu\nu,>} (K) & \simeq G_{JJ,\sym}^{\mu\nu} (K)  \ , \\
G^{>} (Q+K) & \simeq \frac{T}{k^0} \rho(K) \ ,
 \end{align}
to obtain the following expression
\st \Sigma^>_{ab} (Q)= \frac{\tr (T_aT_b)}{F^4} \int \frac{d^4K}{(2\pi)^4}  Q_\mu Q_\nu  G_{JJ,\sym}^{\mu\nu} (K) \frac{T}{k^0} \rho(K) \ . \stp

The soft pion spectral weight takes a simple quasiparticle form
\st \rho(K) \simeq \frac{2\pi}{2v_0 k} \left[ \delta(k^0-v_0k) - \delta(k^0+v_0 k) \right] \ , \stp
with $k=|{\bm k}|$.  The width is small at small $k$,  due to the Goldstone character of the modes, and can be neglected.
Taking the trace
   $\tr (T_a T_b) = \ta \delta_{ab}$ and writing  $\Sigma^>_{ab} (Q) = \delta_{ab} \Sigma^>(Q)$ as in Sec.~\ref{sec:softdetailed}, we find\footnote{
For clarity, the trace of the adjoint representation $\ta$ is kept explicit throughout \Sect{sec:resummedsoft}. In the calculation of \Sect{sec:softdetailed} we have used the scattering amplitude for pions in $SU_V(N)$ with $\ta = N$. This notational difference should be remembered when combining the hard and soft results.} 
% \st
% \Sigma^>(Q) = \frac{\ta}{F^4} \int \frac{d^4K}{(2\pi)^4}  Q_{\mu} Q_{\nu} G_{JJ,\rm sym}^{\mu\nu}(k^0,k) \frac{T}{k^0 }
% \left[ \frac{2\pi}{2 v_0 k} \delta(k^0 - v_0 k ) - \frac{2\pi}{2 v_0 k} \delta(k^0 + v_0 k)  \right] \ .
% \stp
% The integration over $k^0$ gives the evaluation of the symmetric correlator on the light-cone,
\st
\Sigma^>(Q) =  T\frac{\ta}{2} \frac{1}{v_0^2 F^4} \int \frac{d\Omega_k}{4\pi} \int_0^{\infty}  \frac{ dk}{2\pi^2 }   \left[ Q_{\mu} Q_{\nu} \left( G_{JJ,\rm sym}^{\mu\nu}(v_0 k,k)  +  G_{JJ,\rm sym}^{\mu\nu}(-v_0 k,k) \right)   \right] \ . 
\stp

The symmetric correlator $G^{\mu\nu}\equiv G_{JJ,\rm sym}^{\mu\nu}$ is decomposed as
\st
\label{Breakupofgij}
G^{\mu\nu} = \left(\frac{k}{k^0}\right)^2 G_\longi \delta^{\mu}_{0} \delta^{\nu}_{0} + G_\longi \left(\frac{k}{k^0}\right) \left(\delta^{\mu}_{0} \ \hat k^{\nu} +  \delta^{\nu}_{0} \ \hat k^{\mu} \right) 
+ G_\longi  \hat k^{\mu} \hat k^{\nu} +
   G_\trans (\delta^{\mu\nu}_\perp - \hat k^\mu \hat k^\nu) \ ,
\stp
where $\hat k^{\mu} = ( 0, \hat \k)$ and  $\delta^{\mu\nu}_\perp = \delta^{ij}$ for spatial indices,  and zero otherwise.
% \st
% \delta^{\mu\nu}_\perp =  \begin{cases}
%    0  & \mu = 0   \\
%    0  & \nu = 0   \\
%    \delta^{ij} &   \mu,\nu=1,2,3 
% \end{cases} \ .
%    \stp
Finally, this decomposition allows us to write the graph in \Fig{fig:2loopcorrel} as an integral over the longitudinal and transverse parts of the isovector current-current correlator
\st
\label{responseintegralform}
 \Sigma^> (\omega, \q) = T\frac{\ta}{2 \pi} \frac{1}{F^4}
   \int^{\infty}_{-\infty} 
   \frac{ dk}{2\pi }  \left[  \left( \omega^2 + \frac{1}{3}\q^2 \right) G_\longi (v_0 k, k) + \frac{2}{3} \q^2 G_\trans (v_0 k, k) \right] \ .
\stp

\subsubsection{Longitudinal and transverse responses}

Now we will analyze the longitudinal and transverse isovector current-current correlators,
\begin{align}
   I_\longi \equiv& \int^{\infty}_{-\infty} \frac{ dk}{2\pi }   G_\longi (v_0 k, k) \ ,  \\
   I_\trans \equiv& \int^{\infty}_{-\infty} \frac{ dk}{2\pi }   G_\trans (v_0 k, k) \ . 
\end{align}
as they appear in \Eq{responseintegralform}.
For both we are going to exploit the light-cone sum rule presented in Refs.~\cite{Ghiglieri:2013gia,Ghiglieri:2015zma}, derived for the first time in Ref.~\cite{Caron-Huot:2008zna}.  These integrals over the current-current correlation functions evaluate to the isospin susceptibilities up to a contact term. The results are presented in Eqs.~\eqref{eq:IL} and \eqref{eq:IT}.

We will first show how to derive the sum rules for the longitudinal response, by exploiting light-cone causality~\cite{Caron-Huot:2008zna}.
Using the relation between the symmetric and $R/A$ propagators,
\st G_{\longi} \equiv G_{\sym,\longi} (K)= -i \frac{T}{k^0} [G^R_{\longi} (K)- G^A_{\longi} (K) ]  \label{eq:decomp} \stp
we can split $I_\longi$ into retarded and advanced pieces $I_\longi = I^R_\longi + I^A_\longi$, with
\begin{subequations}
   \label{eq:isomomentdefs}
\begin{align}
I_\longi^R & = -i\int_{-\infty}^{\infty} \frac{dk }{2\pi }   \frac{T}{v_0 k} G_\longi^R (v_0 k, k) \ , \label{eq:ILR} \\
I_\longi^A & = i\int_{-\infty}^{\infty} \frac{dk }{2\pi } \frac{T}{v_0 k } G_\longi^A (v_0 k, k) \ . \label{eq:ILA}
\end{align}
\end{subequations}
Note that the apparent pole at the origin is actually ephemeral since at $k\simeq0$  we can use the diffusion expression for the correlator,  Eqs.~(\ref{eq:GRusoftL}) and (\ref{eq:GRusoftT}),  to see that the integrand is finite at the origin. 

\begin{figure}
   \begin{center}
 \includegraphics[width=0.8\textwidth]{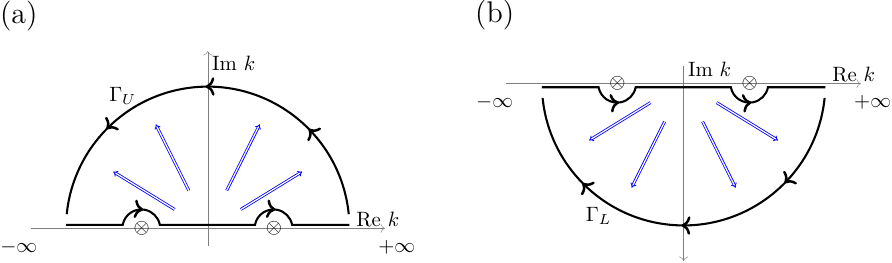}
 \end{center}
   \caption{Contours for the evaluation of functions $I_\longi$ and $I_\trans$: (a) the contour $\Gamma_U$ is for the retarded correlators, which are analytic in the upper complex plane; (b) the contour $\Gamma_L$ is for the advanced correlators, which are analytic in the lower complex plane.} \label{fig:complexplane}
\end{figure}

As the retarded correlator is analytic in the upper complex plane, we can close the integration contour (see Fig.~\ref{fig:complexplane}(a)), and replace the integral by (minus) the one along the upper arc $\Gamma_U$
\st
I_\longi^R = i \int_{\Gamma_U} \frac{dk }{2\pi}    \frac{T}{v_0 k} G^R_\longi (v_0 k, k) \ . \label{eq:ITR2}
\stp
It is understood that the arc at infinity is at the boundary of the soft regime,  $T(T/F)^4 \ll k \ll T$.
To evaluate the retarded correlator on the boundary arc we can use the free kinetic theory as detailed in Eq.~(\ref{eq:Glong}) of App.~\ref{app:correlator},
\st
G^{R}_\longi (K) =\frac{{\cal T}_A}{T} \int_0^{\infty} \frac{dp p^2}{2\pi^2} n_p (1 + n_p)  \int \frac{d\Omega_\p}{4\pi} \frac{k^0 \cos^2 \theta_\p}{-k^0 + v_\p k^z \cos \theta_\p  - i \epsilon}
 \ , \label{eq:GRsoft}
\stp
or by evaluating the hard blue loop in \Fig{fig:2loopcorrel}(a).  Free kinetic theory is appropriate because
on the arc  $k$ is soft compared to the temperature, but  much larger than the collision rate. We note that the isovector susceptibility of
a free gas of pions is  
\st
\label{eq:isospinsusc}
 \chi_I = \frac{{\cal T}_A}{T} \int_0^{\infty} \frac{dp p^2}{2\pi^2} n_p (1 + n_p)  = \ta \frac{T^2}{6}  \, .
\stp
which is reflected in the leading factor of \Eq{eq:GRsoft}.
With $k^0 = v_0 k$ and $v_0, v_\p \simeq 1$, $G^R_\longi(K)$ has a divergence due to cancellation in the denominator. Expanding close to $\cos\theta_\p \simeq 1$, we need to incorporate the collisional width of the hard lines, 
\st
\frac{1}{-v_0 k + v_\p k \cos\theta_\p  -i\epsilon}  \rightarrow \frac{1}{ 
 -v_0 k + v_\p k \cos\theta_\p - i \gamma_p} \simeq \frac{1}{-k + k\cos\theta_p -i\gamma_\p } \ , 
\stp
where we used $k (v_p - v_0) \ll \gamma_p \sim T(T/F)^4$ at the boundary of 
the soft regime.  
Thus on the arc where $k \gg \gamma_p$, the retarded propagator takes the form
\begin{align}
   \label{eq:grlongarc}
   G^{R}_\longi (v_0k,k) & = \frac{{\cal T}_A}{T} \int_0^\infty \frac{dpp^2}{2\pi^2} n_p (1+n_p)  \left[ 1 + \frac12 \log \left(  \frac{i\gamma_p}{2k} \right)   \right] \, .  %\\
   % & = \chi_A \left[ 1 + \frac{i\pi}{4} + \frac12 \log \left( \frac{T}{2k} \right) + \frac{6}{T^3} \int \frac{dpp^2}{2\pi^2} n_p (1+n_p) \frac12  \log \left(  \frac{\gamma_p}{T} \right)  \right]\ , 
\end{align}
% where we factorized the $i \pi/4$ coming from the imaginary unit in $i \gamma_p$. We have also used the definition of $\chi_A$ in Eq.~(\ref{eq:chiAdef}).
The last integration needs to be performed numerically using the function $\gamma_p$ obtained in Sec.~\ref{sec:hardpions}, 
and $G^R_\longi(K)$ on the arc can be conveniently written as
\st
G^{R}_\longi (K) = \chi_I \left[ 1 +  \frac{1}{2} \log \left(  \frac{i\avg}{2k} \right) -0.39 \right]\ ,
\stp
where we introduced the average thermal width $\avg$ of Eq.~(\ref{eq:avthermalgamma}), and the remaining factor is the result of the numerical integration.

Inserting this result into Eq.~(\ref{eq:ITR2}), and parametrizing the arc $\Gamma_U$ as $k = \Lambda e^{i\theta}, \ \theta \in (0,\pi)$ we find 
\st
   I^R_\longi = i T \chi_I \int_0^{\pi}  \frac{i d\theta}{2\pi} \left[  \frac12 \log \left( \frac{\avg}{2 \Lambda} ie^{-i\theta} \right)  +0.61 \right]   = \frac{T \chi_I}{2}  \left[  \frac12 \log \left(  \frac{2 \Lambda}{\avg} \right) -0.61 \right]  \ ,
\stp
where the modulus of $k$ (fixed along the arc) is set equal to $\Lambda$, which serves an UV cutoff for the soft momentum,  regularizing the integral.

For the advanced contribution we repeat the same steps but make use of the lower arc $k = \Lambda e^{-i\theta}$, shown in panel (b) of Fig.~\ref{fig:complexplane}. We obtain the same result as the retarded one, and the two are combined to give the total contribution,
\st
I_\longi = \frac{T\chi_I}{2} \left[\log \left( \frac{\Lambda }{\avg} \right) -0.53 \right] \ . \label{eq:IL}
\stp
One observes that the longitudinal component of the correlation function presents an explicit dependence on $\Lambda$, the scale separating soft and hard domains. Thanks to the light-cone sum rule, only the value of the  propagator at large $k$ is needed when computing the (almost) lightlike integral, and  in this regime the free kinetic approximation in \Eq{eq:GRsoft} is sufficient.

The transverse part is computed analogously, but there is no divergence due to the transverse nature of 
the correlator. Again we write
 $I_\trans= I^R_\trans + I^A_\trans$  as in \Eq{eq:ITR2}, and deform the contours to infinity.
% , with 
% \begin{align}
% I_\trans^R & = -i\int_{-\infty}^{\infty} \frac{dk }{2\pi }   \frac{T}{v_0 k} G_\trans^R (v_0 k, k) \ , \label{eq:ITR} \\
% I_\trans^A & = i\int_{-\infty}^{\infty} \frac{dk }{2\pi } \frac{T}{v_0 k } G_\trans^A (v_0 k, k) \ . \label{eq:ITA}
% \end{align}
The transverse propagator on the arc [analogous to \Eq{eq:GRsoft}] is written in \Eq{eq:Gtrans},
and   on the light cone this reduces reduces to
\st
 G^R_\trans (v_0 k, k)   \simeq - \frac12 \chi_I \int \frac{d\Omega_\p}{4\pi}  (1+ \cos \theta_\p) = - \frac12 \chi_I \ ,  
% \qquad \qquad \textrm{(soft $K$)}
 \label{eq:Grsoft}
\stp
which should be compared to \Eq{eq:IL}. 

The retarded and advanced integrals over the arcs
yield $I^R_\trans = I^A_\trans = T \chi_I/4$, and 
% Then, inserting this simple result into (\ref{eq:ITR}) we get
% \st
% I^R_\trans = i\int_{\Gamma_U} \frac{dk}{2\pi} \frac{T}{v_0k} G^R_\trans (v_0k,k)  \simeq -i \int_0^\pi \frac{i d\theta}{2\pi}   T \chi_A =
%  \frac{1}{4} T\chi_A  \ ,
% \stp
% where the upper arc $\Gamma_U$ has been used. The advanced piece (exploiting the contour $\Gamma_L$) provides the same result.
the sum of the two finally yields  a finite contribution to the self-energy (\ref{responseintegralform}),  
\st I_\trans = \frac12 T\chi_I\ . \label{eq:IT} \stp

\subsubsection{Total contribution in the soft limit}

We have been evaluating \Fig{fig:2loopcorrel}(b), which reflects how the soft pions together with isovector
current response dissipate ultrasoft pion waves. \Eq{responseintegralform} expresses the damping rate in
terms of an integral over the isovector current-current correlator on the light cone;  \Eq{eq:IT} and \Eq{eq:IL} 
determine these integrals as defined in \Eq{eq:isomomentdefs} by exploiting causality. Substituting these expressions, together with isospin susceptibility in \Eq{eq:isospinsusc}, gives the contribution of soft pions to the ultrasoft damping rate
% Substituting \Eq{eq:IT} and \Eq{eq:IL} into \Eq{responseintegralform} with $\omega^2 \simeq \q^2 + m^2$ at lowest order,  we find 
% \begin{align}
% \Sigma^>(Q) & =  \chi_A  \q^2 \frac{\ta T^2 }{2\pi F^4}
% \frac43 \left[ \frac12  \log \left( \frac{\Lambda}{\avg} \right) + \frac12  \log 2 - 0.61 + \frac14 \right]  \nn \\
% & +  \chi_A m^2 \frac{\ta T^2 }{2\pi F^4}
% \left[   \frac12 \log \left( \frac{\Lambda}{\avg} \right) +\frac12 \log 2 - 0.61   \right] \ ,   \qquad (\textrm{soft}) 
% \end{align}
% where we separated explicitly the term carrying the pion mass in the second line.

% Substituting $\chi_A$ and combining the numerical factors, this calculation shows that the coefficient under the logarithm is different for the massless and the massive contribution,
\st
\label{eq:Sigmasoft}
\Sigma^>(Q)   =  \frac{N^2 T^4 }{24\pi F^4} \left\{ (\omega^2 + \tfrac{1}{3} \q^2) \left[ \log\left(\frac{\Lambda}{\avg}\right) -0.53 \right]  + \tfrac{2}{3} \q^2 \right\} \, .
 % \left[  \log \left( \frac{\Lambda}{\avg} \right) + 0.23   \right] 
 % +   m^2 \frac{\ta^2 T^4 }{24\pi F^4}
 % \left[  \log \left( \frac{\Lambda}{\avg} \right) - 0.52  \right] 
% \ . \qquad (\textrm{soft})  \label{eq:Sigmasoft}
\stp
The first and second terms are from the longitudinal and transverse current-current response, respectively. The global factor $\ta^2$ has been replaced by $N^2$ for the $SU_V(N)$ case in the adjoint representation. Thus the $N$ scaling precisely matches with the hard contribution in \Eq{eq:Sigmahard}.

\subsection{Final Results~\label{sec:axialdiff}}

In this section we collect the previous results to provide a final expression for the kinetic coefficients $D_A$ and $D_m$.  Combining the dissipation rates from the hard and soft modes, \Eq{eq:Sigmahard} and \Eq{eq:Sigmasoft} respectively, the cutoff $\Lambda$ cancels, and we find the dissipation rate for ultrasoft pion waves to be
\st
\textrm{Im } \frac{\Sigma^R(\omega,\q)}{\omega} =  \frac{\Sigma^>(\omega, \q)}{2T}   = \frac{N^2 T^3 }{48\pi F^4} \left[ (\omega^2 + \tfrac{1}{3} \q^2) \log\left(\frac{0.86 T}{\avg}\right)  + \tfrac{2}{3} \q^2 \right] \, .
\stp
If the ultrasoft wave is on shell, $\omega^2 = v_0^2 (q^2 + m^2)$,  with $v_0\simeq 1$,  we have
\st
\left. \textrm{Im }  \frac{\Sigma^R(\omega,\q)}{\omega}  \right|_{\omega^2 =v_0^2 (q^2 + m^2)}   =  \frac{N^2 T^3}{36\pi F^4} \log\left( \frac{1.41 T}{\avg}  \right) \q^2  + \frac{N^2 T^3}{48\pi F^4} \log\left(\frac{0.86 T}{\avg}\right) m^2 \, .
\stp 
From this result  we can read off the kinetic coefficients $D_A$ and $D_m$ by comparison with the hydrodynamic expression $\Gamma_q = D_A \q^2 + D_m m^2$, or by the Kubo relations Eqs. \eqref{eq:DA} and \eqref{eq:Dm},  yielding
\begin{align}
   T D_A =&  \frac{N^2 T^4}{36\pi F^4}  \log \left( \frac{1.41T}{\avg}  \right) \, ,  \\
   T D_m =&  \frac{N^2 T^4}{48\pi F^4}  \log \left( \frac{0.86T}{\avg} \right) \, .
\end{align}
Finally, we remind the reader that the coefficients $\sigma_A$ and $\zeta^{(2)}$, introduced in Eqs.~(\ref{eq:dtphi},\ref{eq:JA}), are related to $D_A$ and $D_m$ via \Eqs{eq:sigmarelations}.  Further discussion of the result is given in the next section.

\section{Conclusions and Outlook~\label{sec:outlook}}

As emphasized in the introduction, the chiral limit of $\chi$PT at finite temperature is distinctly different from its massive counterpart, since the Goldstone modes must be included as additional hydrodynamic degrees of freedom~\cite{Son:1999pa,Son:2001ff}. In this limit the long wavelength hydrodynamic effective theory is a kind of non-Abelian superfluid, rather than ordinary hydrodynamics. The goal of this work was to compute the transport
(or kinetic) coefficients of this superfluid theory using $\chi$PT. The relevant Kubo formulas express the kinetic coefficients of this theory in terms of the imaginary part of the pion self-energy $\Sigma^R (\omega, {\bm q})$ for $\omega,{\bm q} \rightarrow 0$. There are only two coefficients: $D_m$, a mass related axial damping coefficient, and $D_A$, the axial charge diffusion coefficient~\cite{Grossi:2020ezz}. A brief summary of the linearized pion effective theory is provided in \Sect{sec:eom}.

As explained in Sec.~\ref{sec:soft}, computing the self-energy with ultrasoft (or hydrodynamic) kinematics, requires a significant resummation. For instance,  $1 \rightarrow 3$ processes, which are normally forbidden by kinematics, turn out to be allowed if one of the external lines is ultrasoft.
This is because the energy violation in the scattering process is small compared to the thermal width of the outgoing hard lines. Similarly, a naive analysis of the scattering rate of soft pions leads to a chiral logarithm, which is (normally) cut off by the mass. However, when the mass is smaller than the inverse mean free path (as is the case in this work)  the natural cutoff is  the  thermal width $\avg$,  and finding the coefficient under the logarithm requires a detailed analysis of the 
 divergence. We were able to analyze the soft sector by exploiting the light-cone sum rules of Caron-Huot\footnote{The technical steps are quite similar to the analysis of the soft fermion contribution to the photon emission rate in hot quark-gluon plasma~\cite{Ghiglieri:2013gia,Ghiglieri:2015zma}.
}~\cite{Caron-Huot:2008zna}.
Our final results for the two kinetic coefficients $T D_A$ and $T D_m$ in $SU_V(N=2)$ are given by 
\begin{align}
TD_A =& \frac{T^4}{9\pi F^4}   \log \left( \frac{1.41 T}{\avg} \right)  \, , \\
T D_m =&  \frac{T^4}{12\pi F^4} \log \left( \frac{0.86T}{\avg} \right)   \ ,
\end{align}
and are of order the typical scattering rate up to the logarithm of the temperature and thermal width. In these expressions,  $F$ is the coefficient in the chiral Lagrangian, \Eq{eq:chilagrangian}, and $\avg = 2.33 \, (T^5/27 F^4)$ is the mean collisional rate for hard pions, \Eq{eq:avthermalgamma}.

The ratio of the two coefficients is
\st r^2 \equiv \frac{D_m}{D_A} = \frac{3}{4} \ ,  \stp
to logarithmic accuracy, and this ratio has a simple interpretation.
Indeed, the factor of $3/4$ is of geometric origin, and reflects how ultrasoft fields with four momentum $Q$
can interact via derivative couplings with a randomly oriented  sample of lightlike particles $v_\p^{\mu} = (1, \hat \p)$ :
\st
 \int \frac{d\Omega_\p}{4\pi}  (v_\p \cdot Q)^2 = \omega^2 +  \frac{\q^2}{3} \simeq m^2 + \frac{4}{3} \q^2 \, .
\stp
The ratio between the $m^2$ and $\q^2$ terms ultimately determines the ratio of $D_m$ to $D_A$ to leading log.

In the limit where the pion mass is much smaller than  thermal width, but still finite, a hierarchy of effective field theories is appropriate. At moderate distances, the superfluid theory of the Goldstone bosons is operative, while at longer distances the system transitions to ordinary hydrodynamics. The physics of the soft pions modes is then reflected in the transport coefficients of the ordinary hydrodynamic theory at long distance. For instance soft pion contribution to the isospin conductivity gives the leading result and reads~\cite{Grossi:2020ezz},
\st
 \sigma_{I}  =    \frac{T}{12 \pi m D_A} \left[   \frac{1+ 2r}{(1 + r)^2 } \right]  .\label{finaltransport3}
\stp
where $r = \sqrt{D_m/D_A} \simeq \sqrt{3/4}$.  
% Note that the result is inversely proportional to 
% the mass, reflecting the almost free nature of the goldstones.
Thus a by-product of our  computation of $D_A$ and $D_m$ is the leading order isospin conductivity at small pion mass.  
Similar superfluid corrections
to the shear and bulk viscosities of ordinary hydrodynamics in terms of $D_A$ and $D_m$ are discussed in \Ref{Grossi:2021gqi}.
Directly computing with isospin conductivity or other corrections from kinetic theory without a sojourn through the superfluid effective theory would be almost hopeless.

We have worked at low temperatures (far from the chiral transition) where the pion velocity is unity up to small corrections. As the system approaches the transition the pion velocity becomes small and the superfluid pion modes become increasingly entangled with the normal modes. It would be interesting to work out the kinetics in this regime even in the large-$N$ limit.

\acknowledgments

We acknowledge Fanglida Yan for early collaboration in the first stages of this research project.  This  work  is  supported  by  the U.S. Department  of  Energy,  Office  of  Science,  Office  of  Nuclear  Physics,  grant  No. DE-FG-02-08ER41450, and by the Deutsche Forschungsgemeinschaft (German Research Foundation) grant numbers 411563442 (Hot Heavy Mesons) and 315477589 - TRR 211 (Strong-interaction matter under extreme conditions).

\appendix

\section{Hard pion self-energy correction~\label{app:hard}}

In this appendix we provide details on the calculation of the hard pion self-energy at finite temperature using $\chi$PT. The calculation is based on several (old) calculations~\cite{Goity:1989gs,Schenk:1993ru}, especially Ref.~\cite{Schenk:1993ru}.
We focus on the self-energy diagram of Fig.~\ref{fig:hardself} for hard pions with typical momentum $p \sim T$. We will first discuss the Wightman self-energy, which is related to the imaginary part of the retarded self-energy discussed next. The Wightman self-energy is 
\begin{multline}
\Sigma^>(P)_{ab} = \frac{\delta_{ab}}{2!} \int \frac{d^3p_2}{(2\pi)^3 2E_2 } 
\frac{d^3p_3}{(2\pi)^3 2E_3 } 
\frac{d^3p_4}{(2\pi)^3 2E_4 } \overline{|\mathcal M|^2} (2\pi)^4\delta^{(4)}(P + P_2 + P_3 + P_4)  \\
\times n_2 (1 + n_3) (1 +n_4)   \ , 
\end{multline}
where the Weinberg amplitude for massless pions with all momentum flowing in is
\begin{multline}
   i\mathcal{M}_{a_1 a_2 a_3a_4}= 
   \frac{i}{F^2} \left[ \delta_{a_1a_2} \delta_{a_3 a_4} (-2 P \cdot P_2)  + \delta_{a_1 a_3} \delta_{a_2 a_4} \left(-2 P \cdot P_3 \right)  + \delta_{a_1 a_4} \delta_{a_2 a_3} \left(-2 P \cdot P_4 \right) \right] \ ,
\end{multline}
and the squared Weinberg amplitude summed over isospin for $a_1,a_2,a_3,a_4$ is given in Eq.~(\ref{eq:Maverage}).

The fluctuation-dissipation theorem relates the emission rate $\Sigma^<(p)/2E_p$ to the absorption rate
\st
  \Sigma^<(p) = e^{-p/T} \Sigma^>(p) \ ,
\stp
and the difference in these rates determines the mean damping rate
\st
\gamma_p \equiv \frac{{\rm Im} \Sigma^R(p)}{p}  = \frac{\Sigma^>(p) - \Sigma^<(p) }{2 E_p} \ .
\stp
The numerical evaluation of $\gamma_p$ is straightforward and the details will not be given.
We  find the following limits at this order
\begin{subequations}
   \label{eq:widthasymptotics}
\begin{align}
   \lim_{p/T \rightarrow 0} \gamma_p =&  \frac{p^2 T^3}{9\pi F^4} \log \left( 1.56 \frac{T}{p} \right)  \, , \\
   \lim_{p/T \rightarrow \infty} \gamma_p= & 
\frac{T^4 p}{F^4}  \left( \frac{\pi}{108} +  0.013 \frac{T}{p}\right)  \ , 
\end{align}
\end{subequations}
where the coefficient under the log in the first expression has been extracted from our numerical data, as has the subasymptotic term at large momenta.  
The mean rate evaluates to
\st
\avg = 2.33 \left( \frac{T^5}{27 F^4} \right) \, ,  \label{eq:dampinghard}
\stp
which is consistent with previous calculations~\cite{Goity:1989gs}.

The real part is significantly more complicated and is treated by Schenk~\cite{Schenk:1993ru}. The thermal correction to the hard on-shell  self-energy  at two-loop order is given by a scattering expansion\footnote{Here and in the remainder of this section all momenta are hard. In the rest of the manuscript $K$ and $Q$ denote soft and ultrasoft momenta respectively. }:
\begin{align}
\Sigma^R (P) & = \Sigma^{R(1)} (P) + \Sigma^{R(2)} (P)\nn \\
& =  \int \frac{d^3q_1}{(2\pi)^3 2 q_1 } n_1 \, T_{\pi\pi}(s)
+ \frac{1}{2} 
\int \frac{d^3q_1}{(2\pi)^3 2 q_1 } 
\int \frac{d^3q_2}{(2\pi)^3 2 q_2 } 
n_1 n_2 \, T_{\pi\pi\pi}^R(P, Q_1, Q_2) \ . 
\end{align}
Here   $T_{\pi\pi}(s)$ is the two pion time-ordered forward scattering amplitude in vacuum, 
and $T_{\pi\pi\pi}^R(P,Q_1, Q_2)$ is the three pion retarded forward scattering  amplitude as  described below.  

The vacuum  scattering amplitude in $\chi$PT to one loop is~\cite{Gasser:1983yg}
\st
T_{\pi\pi}(s) =   \frac{16 s^2 }{F^4} \left[ L^r_1(\mu) +  2  L^r_2(\mu) \right]
  + \frac{5 s^2}{24 \pi^2 F^4} \left[ \log \left( \frac{\mu^2}{s} \right)  +  i \frac{\pi}{2} \right]
  + \frac{2 s^2}{9\pi^2 F^4 }   \, ,
\stp
with $s = -(P+ Q)^2$.  Both $P$ and $Q$ are on shell,  $P^2=Q^2=0$.
The renormalized coupling $L_1^r(\mu)$ and $L_2^r(\mu)$ are given
in \Eq{eq:Lis}, and under a change of renormalization point  $\mu \rightarrow \mu'$ we have
\st
  L_i^r(\mu') = L_i^r(\mu)  - \frac{\gamma_i }{(4\pi)^2 } \log\left( \frac{\mu'}{\mu} \right)
\stp
with $\gamma_1 = 1/12$ and $\gamma_2 = 1/6$. This  leaves the amplitude $T_{\pi\pi}$ unchanged.
Using elementary integrals such as 
\st
  \int \frac{d^3 q_1}{(2\pi)^3 2q_1 } n_1 \, s^2 = \frac{4\pi^2 T^4 p^2}{45}  \, ,
\stp
we find
 \st
 \Sigma^{R,(1)} (P)  = \frac{p^2 T^4}{F^4}  \left\{  \frac{64 \pi^2}{45}  [L^r_1(\mu)+2L^r_2(\mu)]   - \frac{1}{54}\log \left( \frac{Tp}{\mu^2} \right) + i \frac{\pi}{108}-0.022 \right\} \ ,
\stp
The numerical constant can be expressed analytically, but we did not find this useful.
% \begin{multline}
% \Sigma^{R,(1)} (P)  = \frac{p^2 T^4}{F^4}  \left\{  \frac{64 \pi^2}{45}  [L^r_1(\mu)+2L^r_2(\mu)]  - \frac{13}{1620} - \frac{1}{27} \log 2    -   \left( \frac{5}{3}\frac{1}{\pi^4} \zeta'(4) - \frac{6}{324} \gamma_E  \right) \right. \nn \\
%  \left. - \frac{1}{54}\log \left( \frac{Tp}{\mu^2} \right) + i \frac{\pi}{108} \right\} \ ,
% \end{multline} 
% where $\gamma_E$ and $\zeta(z)$ are the Euler-Mascheroni constant  and the Riemann zeta function respectively.
% Numerically 
% \st
% \Sigma^{R,(1)} (P)  = \frac{p^2 T^4}{F^4}  \left\{  \frac{64 \pi^2}{45}  [L^r_1(\mu)+2L^r_2(\mu)]   - \frac{1}{54}\log \left( \frac{Tp}{\mu^2} \right) + i \frac{\pi}{108} -0.0218286 \right\} \ ,
% \stp
The factor $i\pi/108$ determines the asymptotic form of the imaginary part given in \Eq{eq:widthasymptotics}.
% The renormalized LECs 
% \begin{align} 
% \bar{L}_1 &= L_1^r (\mu) - \frac{1}{12(4\pi)^2} \log \left( \frac{M}{\mu} \right) \ , \\
% \bar{L}_2 &= L_2^r (\mu) - \frac{1}{6(4\pi)^2} \log \left( \frac{M}{\mu} \right) \ . 
% \end{align}

The three pion retarded forward scattering amplitude $T_{\pi\pi\pi}^R(P_1, P_2, P_3)$ takes
the form~\cite{Schenk:1993ru} 
\begin{multline}
 T^R_{\pi\pi\pi}(P_1, P_2, P_3) = 
 \frac{{\mathcal V}(P_1, P_2, P_3) }{(P_1 + P_2 + P_3)^2 }
  + \frac{{\mathcal V}(P_1, -P_2, P_3) }{(P_1 - P_2 + P_3)^2 }
  + \frac{{\mathcal V}(P_1, P_2, -P_3) }{(P_1 + P_2 - P_3)^2 } \\
  + \frac{{\mathcal V}(P_1, -P_2, -P_3) }{(P_1 - P_2 - P_3)^2 } \, .
  \end{multline}
Here $P_1^2=P_2^2=P_3^2=0$ and $p^0_i = |{\bm p}_i|$.  
The vertex structure  is
\st {\mathcal V}(P_1, P_2, P_3)= \frac{2}{3F^4} [((P_0-P_1)\cdot (P_2-P_3))^2+((P_0-P_2)\cdot(P_1-P_3))^2+((P_1-P_2) \cdot (P_0-P_3))^2] \ , 
\stp
where $P_0 + P_1 + P_2 + P_3=0$.
It is understood that the timelike component in these expressions is retarded, e.g.
\st
 \frac{1}{(P_1 + P_2 + P_3)^2 } = \frac{1}{-(p_1^0 + p_2^0 + p_3^0 + i\epsilon)^2  + (\p_1 +\p_2 + \p_3)^2 } \ .
\stp
The three pion scattering amplitude contribution to $\Sigma^R$ can be compactly written
\begin{multline}
\Sigma^{R,(2)}(p,p)  = \frac{1}{2} \int \frac{d^4 Q_1}{(2\pi)^4}  \frac{d^4 Q_2}{(2\pi)^4}  \frac{d^4 Q_3}{(2\pi)^4} (2\pi)^4 \delta^{(4)}(P+Q_1+Q_2+Q_3) {\mathcal V}(P,Q_1,Q_2,Q_3) \nn \\
 \times 2\pi \delta(Q_1^2) 2\pi \delta(Q_2^2) n_{1} n_{2}
 \frac{1}{Q^2_3} \ , 
\end{multline}
where  $n_{1} = n(|{\bm q}_1|)$. 

The numerical integration is not completely straightforward, and therefore we will indicate the steps when $q_1^0=|{\bm q}_1|$ and $q_2^0=|{\bm q}_2|$, leaving the case when $q_1^0=-|{\bm q}_1|$ to the reader. The goal is to choose a
coordinate system, which is well adapted to the singular denominator $Q_3$.
The phase space is 
\st
   \int_{PS} =  % \int \frac{d\Omega_\p}{4\pi} 
   \int_1 \int_2 \int_3 
   2\pi \delta_{+}(Q_1^2)
   2\pi \delta_{+} (Q_2^2) 
(2\pi)^4 \delta(P + Q_1 + Q_2 + Q_3)  \\
\stp
where $\int_1 = \int d^4Q_1/(2\pi)^4$, and $\delta_+(P^2) =\theta(p^0) \delta(P^2)$.  We define 
   ${\bm k_1}=\p +  {\bm q}_1$ and ${\bm k_2} = \k_1 + {\bm q}_2$
so that  
\st
  \frac{1}{Q_3^2} =  \frac{1}{-(p + q_1 + q_2 + i\epsilon)^2 + k_2^2 }
\stp
 We take $\p$ on the $z$ axis, and use the azimuthal invariance to place ${\bm q}_1$  in the $xz$ plane.
% respect to $\hat \p$ by exploiting 
% the azimuthal  invariance of the self-energy, 
% and
We then parametrize the angle between $\p$ and ${\bm q}_1$ with the magnitude of $k_1 =|\k_1|$, 
\st
\frac{1}{2} \int_{-1}^{1} d\cos \theta_{pq_1}  =  \int_{|p - q_1|}^{p + q_1} \frac{k_1 dk_1}{2 p q_1} \, .
\stp
Similarly,  we parametrize the angles of ${\bm q}_2$ with respect to the
 $\k_1$ axis, i.e.  $\cos\theta_{k_1 q_2}$ and $\phi_2$,  by the magnitude of $k_2=|{\bm k}_2|$ 
\st
\frac{1}{2} \int_{-1}^{1} d\cos \theta_{k_1q_2}  = 
\int_{|k_1 - q_2|}^{k_1 + q_2} \frac{k_2 dk_2}{2 k_1 q_2} \ .
\stp
With this parametrization we find
\begin{align}
\int_{PS} = \frac{1}{64 \pi^4 p} \int_0^{\infty}  dq_1  
\int_0^{\infty} dq_2  \int_{|p - q_1|}^{p + q_1} dk_1
\int_{|k_1 - q_2|}^{k_1 + q_2} k_2 dk_2 \int \frac{d\phi_2}{2\pi} \, .
\end{align}
The advantage of this parametrization is that the integral over $k_2$ and $\phi_2$ can be done analytically (with computer algebra) giving a logarithmic dependence.  This logarithm is evaluated correctly by including a finite $i\epsilon$ and using complex arithmetic with $\epsilon \sim 10^{-10}$. 
The remaining integrals over $q_1$, $q_2$, and $k_1$ are done via Monte Carlo integration, which gracefully handles the sharp boundaries and logarithmic singularities imposed by the $i\epsilon$ prescription. 
Ultimately all of the vectors are explicitly needed to evaluate ${\mathcal V}$.  If $\k_1$ is taken along the $Z$ axis, then we have the vectors
\begin{subequations}
\begin{align}
   {\bm p} =& (p \sin\theta_{k_1 p}, \, 0, \,  p \cos\theta_{k_1 p} ) \, , \\
   {\bm q}_1 =& (q_1 \sin\theta_{k_1 q_1}, \, 0, \, q_1 \cos\theta_{k_1 q_1} ) \, , \\
   {\bm q}_2 =& (q_2 \sin\theta_{k_1 q_2} \cos\phi_2,\, q_2 \sin\theta_{k_1 q_2} \sin\phi_2, \,  q_2 \cos\theta_{k_1 q_2} ) \, ,
  \end{align}
\end{subequations}
and the angles can be worked out, e.g. $\cos\theta_{k_1p} = (k^2_1 + p^2 - q_1^2)/2p k_1$.  .
% Some angles which are needed for evaluating the integrand are given below
% \begin{subequations}
% \begin{align}
%  \cos\theta_{pq_1} =  \frac{1}{2 p q_1} (k_1^2  - p^2 - q_1^2) \, , \\
%  \cos\theta_{k_1q_2} = \frac{1}{2 k_1 q_2} (k_2^2  - k_1^2 - q_2^2) \, , \\
%  \cos\theta_{k_1q_1}  = \frac{1}{2 q_1 k_1} (k_1^2  - p^2 + q_1^2) \, ,  \\
% \cos\theta_{k_1p} = \frac{1}{2 p k_1} (k^2_1 + p^2 - q_1^2) \, .
%  \end{align}
%  \end{subequations}

When $q_1^0 = -|{\bm q}_1|$ and $q_2^0=|{\bm q}_2|$ a slightly modified parameterization is necessary. In this case we define $\k_1 = \p - {\bm q}_1$ and $\k_2 = \k_1 + {\bm q}_2$, but otherwise make similar steps.
A very good check of the integration procedure is 
that the imaginary part, which is computed in the same go as the real part,
is in agreement with the elementary evaluation described above. 
In addition, our numerical results for $v_0$ are in agreement by Toublan who used a completely different parametrization~\cite{Toublan:1997rr}.

% In Eq.~(\ref{eq:avthermalgamma}) we have defined mean scattering rate,
% \st
% \avg = \frac{\int d^3p \ n_p \gamma_p }{\int d^3p \ n_p} \ ,
% \stp
% which at 1-loop level takes an exact expression,
% \st
% \gamma_p = \frac{T^5}{F^4} \frac{\pi p}{108T} \ \rightarrow  \quad \avg =  \frac{\pi^5}{120 \zeta(3)} \left( \frac{T^5}{27 F^4} \right)  \qquad (\textrm{1-loop}) \ .
% \stp
% At 2-loop order only a numerical solution for $\gamma_p$ is possible. It is shown in Fig.~\ref{eq:avthermalgamma}. 
% We also find the following limits at this order
% \st 
% \lim_{p/T \rightarrow 0} \gamma_p = \frac{T^5}{F^4} \frac{p}{9\pi T} \log \left( 1.56 \frac{T}{p} \right) \quad ; \quad \lim_{p/T \rightarrow \infty} \gamma_p= 
% \frac{T^5}{F^4} \frac{p}{T} \left( \frac{\pi}{108} +  0.013 \frac{T}{p}\right)  \ , \stp
% and the mean rate
% \st
% \avg = 2.33 \left( \frac{T^5}{27 F^4} \right) \qquad (\textrm{2-loop}) \label{eq:dampinghard}
% \stp
% which is consistent with previous calculations~\cite{Goity:1989gs}.

The real part of the self-energy determines the velocity
\st 
v^2(p) = 1- \frac{\textrm{Re } \Sigma^R (p)}{p^2} \ . 
\stp
The value at $p\rightarrow 0$ gives,
\st 
v_0^2 = v^2(p=0)= 1 - \frac{T^4}{27 F^4} \log \left( \frac{\Lambda_\Delta}{T} \right) \ , \stp
with
\st 
\log \left( \frac{\Lambda_\Delta}{T} \right) = 
- \log \left( \frac{T}{\mu}  \right)+
\frac{192\pi^2}{5}[L^r_1(\mu)+2L^r_2(\mu)] +0.54 \ . 
\stp
The numerical result for $v(p)-v_0$ for all values of momenta are plotted in Fig.~\ref{fig:quasiparticles}.

\section{Off-shell regularization of the ultrasoft pion self-energy}
\label{app:soft}

In this appendix we review the calculation of the two-loop self-energy with an ultrasoft external momentum $Q$, introducing  corrections due to collisional widths to the previous result of Ref.~\cite{Smilga:1996cm}, which are valid to logarithmic accuracy. 

In the derivation initiated in Sec.~\ref{sec:softdetailed} we arrived to the expression~(\ref{eq:sigmaaux}) for $\Sigma^>(Q)$, where the Wightman functions appearing in the integrand incorporate the thermal widths in a quasiparticle approximation. These are detailed in Eq.~(\ref{eq:Gpi}).
To simplify those expressions we introduce parallel and perpendicular momenta, $\p_i = (p_{\pl,i}, \p_{\perp,i})$ according to a preferred direction marked by $\p_2$.  Then, conservation of momentum reads
\st  p^\pl_2 + p^\pl_3 + p^\pl_4 =  0  \ ,  \qquad   \p^\perp_3 + \p^\perp_4  = 0 \ . \label{eq:momconser} \stp

After a trivial integration of $p_4^0$, the integrations in $p^0_2,p^0_3$ can also be made explicitly for small $\gamma_i$. Notice that in the numerator we have a term that depends on $p_2^0$: $(-2 Q\cdot P_2)^2 = 4 (\omega p_2^0 - q^\pl p_2^\pl)^2$. We use
\begin{multline}
\int  \frac{dp_2^0}{2\pi}  \frac{dp_3^0 }{2\pi}
\frac{s_2 s_3 s_4}{2E_2 2 E_3 2 E_4}  
\ 4 (\omega p_2^0 -q^\pl p_2^\pl)^2 \\
\frac{\gamma_2 }{(p_2^0 - s_2 E_2)^2 + (\gamma_2/2)^2  } 
\frac{\gamma_3 }{(p_3^0 - s_3 E_3)^2 + (\gamma_3/2)^2  } 
\frac{\gamma_4 }{(-p_2^0 - p_3^0 - s_4 E_4)^2 + (\gamma_4/2)^2  } \\
= 4 (\omega s_3 E_3 - q^\pl p_2^\pl )^2 \
\frac{s_2 s_3 s_4}{2 E_2 2 E_3 2 E_4}   
\frac{ (\gamma_2 + \gamma_3  + \gamma_4)  }{ (s_2 E_2 + s_3 E_3 + s_4 E_4)^2 + (\gamma_2 + \gamma_3  + \gamma_4)^2/4 } + {\cal O}(\gamma^2) \label{eq:intzero}
\end{multline}

With this result, the pion self-energy is seen to be dominated by the kinematic regime where 
\st
s_2 E_2 + s_3 E_3 + s_4 E_4  \sim \gamma \ ,
\stp
where $\gamma \equiv \gamma_2+\gamma_3+\gamma_4$. The integration over remaining momenta can be performed taking into account the appropriate signs ($s$ factor) appearing in Eq.~(\ref{eq:intzero}). One can use
\st
s_i E_i = v_{p,i} p^\pl_i  + \frac{p_{\perp,i}^2}{2p_i^\pl}
\stp
to write the combination
\begin{align}
   s_2 E_2 + s_3 E_3 + s_4 E_4 =& v_2 p^\pl_2 + v_3 p_3^{\pl} + v_4 p_4^\pl + \frac{p_{\perp,3}^2 }{2 p^\pl_3} + \frac{p_{\perp,4}^2}{2 p^\pl_4} = \sum_{i=2,3,4} v_i p^\pl_i  - \frac{p_2^\pl p_{\perp,3}^2 }{2 p^\pl_3 p^\pl_4}  \ ,
\end{align}
where Eq.~(\ref{eq:momconser}) has been used.

\begin{figure}[ht]
  \centering
  \includegraphics[width=150mm]{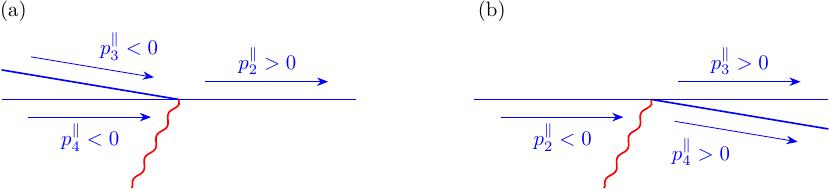}
  \caption{Joining (a) and splitting (b) effective $1\leftrightarrow 3$ processes which can take place with the kinematic constraints of the ultrasoft pion self-energy.}
  \label{fig:splitjoin}
\end{figure}

Denoting by $\theta$ the angle between $\bm{q}$ and $\p_2$, and $\Omega_2$ the solid angle subtended by $\p_2$ we find
\begin{multline}
   \Sigma^>(Q)  = \frac{N^2}{4} \frac{2}{F^4}    \frac{1}{2!} 
   \frac{1}{\pi} \int \frac{ d\Omega_2}{4 \pi} \int_{-\infty}^{\infty} \frac{dp_2^\pl}{2\pi} p^{\pl,2}_2
   \int_{-\infty}^{\infty} \frac{dp_3^\pl}{2\pi} 
   \int \frac{d^2\p_{\perp,3}}{(2\pi)^2 }
   \int_{-\infty}^{\infty} \frac{dp_4^\pl}{2\pi} \int \frac{d^2\p_{\perp,4}}{(2\pi)^2 }  \\
 \times   \frac{ 1 + n(p_2^\pl)}{2 p_2^\pl}    \frac{ 1 + n(p_3^\pl)}{2 p_3^\pl}  
  \frac{ 1 + n(p_4^\pl)}{2 p_4^\pl}  \ 4 (p_2^\pl)^2  (\omega - \q \cos\theta)^2  \\
\times (2\pi)^2    \delta^{(2)} ( \p_{\perp,3}+\p_{\perp,4} ) \ 2\pi \delta(p^\pl_2+p^\pl_3+p_4^\pl) \frac{\gamma}{ \left( \sum_{i} v_i p_i^\pl - \frac{p_2^\pl p^2_{\perp,3}}{2p_3^\pl p_4^\pl} \right)^2 + \gamma^2/4} \ ,
\end{multline}
The integral over $\p_{\perp,4}$ is trivially performed. Now we focus on the term in the denominator,
\st
\left( \sum_i v_i p_i^\pl  - \frac{p_2^{\pl}}{2 p_3^\pl p_4^\pl} p_{\perp,3}^2 \right)^2
= 
\left( s_{234} \sum_i v_i p_i^\pl  - \left|\frac{p_2^{\pl}}{2 p_3^\pl p_4^\pl} \right| p_{\perp,3}^2 \right)^2 \ ,
\stp
where $s_{234}= {\rm sign }(p^\pl_2 p^\pl_3 p^\pl_4)$ can take two values depending on the physical kinematic process,
\st s_{234}= 
\begin{cases}
   +1  &  {\rm joining} \quad (p^\pl_i,p^\pl_j<0;p^\pl_k>0)  \quad \ ,  \\
   -1  &  {\rm splitting} \quad (p^\pl_i,p^\pl_j>0;p^\pl_k<0)  \quad \ ,
\end{cases}  \label{eq:s234}
\stp
where $i,j,k$ label the three different pions. In the joining process two incoming pions (with negative momentum) produce a single final pion (with positive momentum), whereas in the splitting process a single incoming pion gives two outgoing pions. These processes are depicted in Fig.~\ref{fig:splitjoin} for the case where the isolated pion is $p_2^\pl$.

In both cases one has
\st
s_{234} \sum_i v_i p_i^{\pl} = E_{a + b}  -  E_{a} - E_{b}  \equiv \delta E(p_a, p_b)
\stp
where $p_a $ and $p_b$ are the two smaller momenta. It is computed from the real part of the self-energy correction in Sec.~\ref{sec:hardpions},
\st  \label{eq:deltaE}
\delta E = E_{a +b} -  E_{a}  - E_{b} = (v_{a+b} - v_{a}) p_a  + (v_{a+b} - v_{b} ) p_b > 0  \ ,
\stp
where $v_a$ are the velocities of the pions. We remind the reader that $\delta E$ is soft, and therefore the same order as $\gamma$.

Doing a change of variables one can perform the integration over $\p_{\perp,3}$. Defining $u \equiv \left|\frac{p_2^{\pl}}{2 p_3^\pl p_4^\pl}\right| p^2_{\perp,3}$, the pion self-energy reads
\begin{multline}
   \Sigma^>(Q) = \frac{N^2}{F^4}    \frac{1}{\pi} \int \frac{ d\Omega_2}{4 \pi} (\omega - \q \cos \theta)^2 \int_{-\infty}^{\infty} \frac{dp_{3}^\pl}{2\pi} \int_{-\infty}^{\infty} \frac{dp_4^{\pl}}{2\pi} \int_{-\infty}^{\infty} \frac{dp_2^\pl}{2\pi}   p^{\pl,2}_2  \ 2\pi \delta(p^\pl_2+p^\pl_3+p_4^\pl)  \\
\times  \frac{ 1 + n(p_2^\pl)}{2 p_2^\pl}  
    \frac{ 1 + n(p_3^\pl)}{2 p_3^\pl}  
  \frac{ 1 + n(p_4^\pl)}{2 p_4^\pl}   \left| p^\pl_2 p^\pl_3 p^\pl_4 \right|  \int_0^\infty \frac{du}{2\pi}   \frac{\gamma}{(\delta E - u)^2 +\gamma^2/4} \ ,
\end{multline}
where the last integral gives
\st
\int_0^{\infty} \frac{du}{2\pi}   \frac{\gamma }{ (u -\delta E)^2  + \gamma^2/4 }
=   \frac{1}{2} +  \frac{1}{\pi} \tan^{-1} \left( \frac{2 \delta E}{\gamma} \right) \ . \label{eq:intdamp}
\stp

Performing also the trivial angular integration we find  
\begin{multline}
   \Sigma^>(Q)  =  \frac{N^2}{16\pi F^4 } \big(\omega^2 + \tfrac{1}{3} \q^2\big) \int_{-\infty}^{\infty} \frac{dp^\pl_3}{2\pi}   \int_{-\infty}^{\infty} \frac{dp^\pl_4}{2\pi}  \int_{-\infty}^{\infty} \frac{dp^\pl_2}{2\pi}   p^{\pl,2}_2 \ 2\pi \delta(p^\pl_2+p^\pl_3+p^\pl_4)  \nn \\
 \times  (1 + n_2)    (1 + n_3)    (1 + n_4) s_{234} 
      \ 2  \left[\frac{1}{2} +  \frac{1}{\pi} \tan^{-1} \left(  \frac{2\delta E}{\gamma} \right) \right] \ , 
      \end{multline}
where we called $n_i \equiv n(p^\pl_i)$.

We can now consider explicitly the cases where each of the three momenta are maximal. First one distinguishes the case in which $p_2^\pl$ is maximal with either sign---i.e., positive for joining, and negative for splitting processes. Calling this contribution the $s$-channel, we eventually find
\begin{multline}
   \left. \Sigma^>(Q) \right|_{s}  =  \frac{N^2}{4F^4} \big(\omega^2 + \tfrac{1}{3} \q^2\big) 
   \int_0^{\infty} \frac{p^{\pl,2}_2 dp^\pl_2}{2\pi^2} 
   \int_0^{p^\pl_2/2} \frac{dp^\pl_3}{2\pi} 
   n_2 (1 + n_3) (1 + n_4)  \\ 
   \times 2 \left[ \frac{1}{2} +  \frac{1}{\pi} \tan^{-1} \left( \frac{2\delta E}{\gamma} \right) \right] \ , \label{eq:finalsigmas}
   \end{multline}
and after performing the final integrations  obtain
\st \left. \Sigma^>(Q) \right|_{s} = \frac{N^2 T^4}{48\pi F^4} \big(\omega^2 + \tfrac{1}{3} \q^2\big) 
 \left[  \log \left( \frac{T}{\Lambda} \right) + 0.63 +0.34 \right] \ . \stp
Here the first term gives the IR divergence behavior, the second term gives the finite coefficient of the divergent term, and the last term the finite contribution of the $\tan^{-1}$ term. 

When $p_3^\pl$ or $p_4^\pl$ are maximal momenta (with either sign), the contributions are similar. We call these $t$ and $u$ contributions, respectively. Combining these two, we find 
\begin{multline}
   \left. \Sigma^>(Q) \right|_{t+u}  = \frac{N^2}{4F^4}  \big(\omega^2 + \tfrac{1}{3} \q^2\big) 
   \int_0^{\infty} \frac{dp^\pl_3}{2\pi} 
   \int_0^{p^\pl_3} \frac{p_2^{^\pl,2} dp^\pl_2 }{2\pi^2}
   n_3 (1 + n_2) (1 + n_4) \\
   \times 2 \left[ \frac{1}{2} +  \frac{1}{\pi} \tan^{-1} \left( \frac{2 \delta E}{\gamma} \right) \right] \ , \label{eq:finalsigmatu}
   \end{multline}
which finally gives
\st 
\left. \Sigma^>(Q) \right|_{t+u} =  \frac{N^2 T^4}{48\pi F^4}  \big(\omega^2 + \tfrac{1}{3} \q^2\big) \left[  \log \left( \frac{T}{\Lambda} \right) -0.46 +0.23 \right] \ . \stp
The total contribution from the internal hard pions reads
\st \Sigma^>(Q) = \left. \Sigma^>(Q) \right|_s + \left. \Sigma^>(Q) \right|_{t+u}= \frac{N^2 T^4}{24\pi F^4}  \big(\omega^2 + \tfrac{1}{3} \q^2\big) \left[  \log \left( \frac{T}{\Lambda} \right) + 0.37 \right] \ , \label{eq:Sigmagreater} \stp
where the coefficient of the logarithm coincides with the result quoted in Ref.~\cite{Smilga:1996cm}. In addition, we provide the complete result for the coefficient under the logarithm.

From this result, it is clear that a kinetic treatment in the soft sector is then needed to match the IR logarithmic divergence of (\ref{eq:Sigmagreater}) caused by the softening of the internal pion momentum. 
If the pion mass had been taken to be large compared to $T(T/F)^4$ ---like in the usual $\chi$PT power counting---then the mass would directly regulate the IR divergence. However,  here the mass is ultrasoft and we need to carefully analyze the soft sector to regulate the divergence with the quasiparticle width.

\section{Soft and ultrasoft isovector current-current correlators~\label{app:correlator}}

We consider the retarded/advanced current-current correlation function,
\st 
G_{JJ}^{R/A,\mu\nu} (K) =  \pm i\int d^4X e^{iK\cdot (X-Y)} \llangle [ J^\mu(X), J^\nu (Y) ]\rrangle \theta(\pm(X^0-Y^0)) \ .
 \stp

Let us first consider the momentum $K$ as soft, which is related to the kinetic scale.
For high momentum, $T(T/F)^4 \ll K \ll T$, the correlator can be obtained via  kinetic theory. When collisions are not yet relevant for such modes a hard-thermal-loop approach is possible~\cite{Manuel:1997zk,Laine:2016hma}. The spatial components read
\st G^{R/A,ij} (K)= \frac{{\cal T}_A}{T} \int_0^{\infty} \frac{dp p^2}{2\pi^2} n_p (1 + n_p) 
\int \frac{d\Omega_\p}{4\pi} \frac{k^0 v_\p^i v_\p^j}{v_\p \cdot K \mp i \epsilon} \ , 
\stp
where $v_\p^\mu=(1,\bm{v}_\p)$ is light-like. Incidentally, if $\bm{v}_\p$ does not depend on $p$, then one can identify the
isovector susceptibility $\chi_I$, defined in Eq.~(\ref{eq:isospinsusc}), and write
\st G^{R/A,ij} (K)= \chi_I \int \frac{d\Omega_\p}{4\pi} \frac{k^0 v_\p^i v_\p^j}{v_\p \cdot K \mp i \epsilon} \ . \qquad \qquad \textrm{(soft $K$)} 
\stp

In the general case, one can identify the transverse and longitudinal components,
\begin{align} 
    G^{R/A}_\trans(K) & = \frac12 (\delta_{ij} - \hat{k}_i \hat{k}_j) G^{R/A,ij} (K) \label{eq:Gtrans}  \\
  &=  \frac{{\cal T}_A}{2T} \int_0^{\infty} \frac{dp p^2}{2\pi^2} n_p (1 + n_p)  \int \frac{d\Omega_\p}{4\pi} \frac{k^0 (1- \cos^2 \theta_\p)}{-k^0 + v_\p k^z \cos \theta_\p  \mp i \epsilon}  \ , \nn \\
  G^{R/A}_\longi (K) & = \hat{k}_i \hat{k}_j G^{R/A,ij} (K) \label{eq:Glong}  \\
  &= \frac{{\cal T}_A}{T} \int_0^{\infty} \frac{dp p^2}{2\pi^2} n_p (1 + n_p)  \int \frac{d\Omega_\p}{4\pi} \frac{k^0 \cos^2 \theta_\p}{-k^0 + v_\p k^z \cos \theta_\p  \mp i \epsilon} \ ,  \nn 
\end{align}
where we have chosen the vector $\k$ pointing along the $Z$ axis, and $\hat{k}^i=k^i/|\k|$.

When the external $K$ is ultrasoft, then one needs to evaluate the correlation function in the full hydrodynamic limit, incorporating all the effects of collisions. In this limit the retarded correlator is~\cite{forster2018hydrodynamic,Kovtun:2012rj}
\begin{align}
G^{R,00} (k^0,\k) & = \frac{\chi_I D_I |\k|^2}{-ik^0 + D_I |\k|^2 } \ , &\textrm{(ultrasoft $K$)} \\
G^{R,0i} (k^0,\k)& = \frac{\chi_I D_I k^0 |\k|}{-ik^0+D_I |\k|^2} \hat{k}^i \ , & \textrm{(ultrasoft $K$)}  \\
G^{R,ij} (k^0,\k)& = (\delta^{ij}-\hat{k}^i \hat{k}^j) i \chi_I D_I k^0 +\frac{\chi_I D_I k^{0,2}}{-ik^0+D_I|\k|^2} \hat{k}^i \hat{k}^j \ , & \textrm{(ultrasoft $K$)} 
\end{align}
where $D_I$ is the isovector diffusion coefficient. From here we can obtain in particular,
\begin{align} 
G^R_\longi (k^0, \k) & = \hat{k}^i \hat{k}^j G^{R,ij} (k^0,\k) = 
\frac{\chi_I D_I k^{0,2}}{-ik^0 + D_I |\k|^2} \ , \qquad \qquad \textrm{(ultrasoft $K$)}\label{eq:GRusoftL} \\
G^R_\trans (k^0,\k)  & =  \frac12 (\delta^{ij} - \hat{k}^i \hat{k}^j) G^{R,ij} (K)  = i\chi_I D_I k^0 \ .  \qquad \qquad \textrm{(ultrasoft $K$)}  \label{eq:GRusoftT} 
\end{align}

\section{Goldstone boson scattering in $SU(N)$\label{app:sun}}
In Eq.~(\ref{eq:Maverage}) we have presented the well-known formula for the pion-pion average scattering amplitude square for $N=2$ flavors in the massless case~\cite{Weinberg:1966kf,Gasser:1983yg}. In this appendix we generalize this result for massive pions living in the adjoint representation of $SU_V(N)$ with $N \ge 2$. The scattering amplitude $i {\cal M}_{a_1 a_2,a_3 a_4}$ (with indices $a_i = 1,..., N^2-1$) is taken from the results of Ref.~\cite{Chivukula:1992gi}. For arbitrary $N$, several structures appear,
\begin{align} 
{\cal M}_{a_1 a_2, a_3a_4} & = \delta_{a_1 a_2} \delta_{a_3 a_4} A (s,t,u) + \delta_{a_1 a_3} \delta_{a_2 a_4} A(t,s,u) +\delta_{a_1 a_4} \delta_{a_2 a_3} A(u,t,s)  \label{eq:iMSUN} \\
& + d_{a_1 a_2 b} d_{a_3 a_4 b} B(s,t,u) + d_{a_1 a_3 b} d_{a_2 a_4 b} B(t,s,u) + d_{a_1 a_4 b} d_{a_2 a_3 b} B(u,t,s) \ , \nn
\end{align}
where only the first line survives in the usual $N=2$ case. The two scalar functions, $A(s,t,u)$ and $B(s,t,u)$ are obtained in $\chi$PT at leading order,
\begin{align}
 A (s,t,u) & = \frac{2}{N} \frac{s-m^2}{F^2} \ , \\
 B (s,t,u) & = \frac{s-m^2}{F^2} \ . 
\end{align}
In Eq.~(\ref{eq:iMSUN}) $d_{abc}$ are the totally symmetric $d$-symbols of $SU(N)$, and a sum over repeated indices is understood.
The average scattering amplitude square is defined in Eq.~(\ref{eq:Maverage}) ,
\st \overline{|\mathcal M|^2} = \frac{1}{N^2-1} \sum_{a_1a_2a_3a_4} | \mathcal M^{a_1a_2}_{a_3a_4}|^2 \ ,
\stp
which is normalized by the factor $d_A=N^2-1$. Replacing the scattering amplitudes, and after a series of algebraic steps, we arrive to the simplified result
\st \overline{|{\cal M}|^2} = N^2 \ \frac{s^2+t^2+u^2}{2F^4} - 2 \ \frac{N^4+2N^2-6}{N^2} \ \frac{m^4}{F^4} \ . \label{eq:avMSUN}
\stp
Notice that the mass-independent term gets a simple $N^2$ factor, while the term proportional to $m^4$ carries a nontrivial dependence with $N$. We stress that---as explained in Sec.~\ref{sec:softdetailed}---this second term plays no role in the ultrasoft pion self-energy at the order we are working on, as the leading mass dependence ${\cal O} (m^2)$ comes from the  dispersion relation of the external pion. 
Finally, notice that for $N=2$ the expression in Eq.~(\ref{eq:avMSUN}) reduces to the well-known result (\ref{eq:M2SU2}) for the pion-pion scattering in $SU(2)$.

\bibliography{refs}

\end{document}